\documentclass[journal]{IEEEtran}
\usepackage{soul}
\usepackage{cite}

\ifCLASSINFOpdf
  \usepackage[pdftex]{graphicx}
  \graphicspath{{./figures/}}
  \DeclareGraphicsExtensions{.pdf,.jpeg,.png}
\else
\fi
\usepackage{multirow}

\usepackage{amsmath}
\usepackage{array}

\ifCLASSOPTIONcompsoc
 \usepackage[caption=false,font=normalsize,labelfont=sf,textfont=sf]{subfig}
\else
 \usepackage[caption=false,font=footnotesize]{subfig}
\fi

\usepackage{color}
\begin{document}
%
\title{Consensus Control for Coordinating Grid-Forming and Grid-Following Inverters in Microgrids}
%
%
%

\author{Ankit~Singhal,~\IEEEmembership{Member,~IEEE,}
        ~Thanh~Long~Vu$^*$,~\IEEEmembership{Member,~IEEE,}
        and~Wei~Du,~\IEEEmembership{Member,~IEEE,}
 \thanks{The authors are with Pacific Northwest National Laboratory, Richland, WA 99352.
 $^*$ Corresponding author: Thanh Long Vu (email: thanhlong.vu@pnnl.gov).}}
\maketitle
\begin{abstract}
In a purely inverter-based microgrid, both grid-forming (GFM) and grid-following (GFL) inverters will have a crucial role to play in frequency/voltage regulation and maintaining power sharing through their grid support capabilities. Consequently, the coordination between these two technologies becomes increasingly important for optimal system performance. However, the existing work does not consider GFL's potential to participate in a secondary control in coordination with GFM, thus not able to utilize the full capability of inverter resources. In this paper, we show that it is possible to fully coordinate the GFL and GFM inverters to achieve accurate power sharing, frequency/voltage regulation, and circulating var mitigation in networked microgrids even without the support of any synchronous generators or the bulk power system. We use the leader-follower consensus framework to develop a GFM-GFL coordination control. The effectiveness of the proposed coordination is verified under different disturbances and communication degradation. In addition, we find that the proposed fully-coordinated secondary control outperforms other approaches such as un-coordinated and partially-coordinated secondary controls, in aspects of load sharing and frequency and voltage regulation. Overall, this study emphasizes the need and benefits of GFL-GFM coordination in microgrids.
\end{abstract}

\begin{IEEEkeywords}
Inverters, DER, grid-forming, grid-following, consensus, frequency stability, voltage regulation, power sharing.
\end{IEEEkeywords}
\vspace{-3mm}
%
\IEEEpeerreviewmaketitle
\section{Introduction}
\vspace{-0.5mm}

A microgrid is defined as a group of interconnected loads and distributed energy resources (DERs) within clearly defined electrical boundaries that acts as a single controllable entity and can operate in grid-connected or islanded mode. \cite{Lasseter2001,TON201284}. A microgrid has the potential to provide energy surety to critical services, improve the reliability of power grids during extreme events, and can evolve as a key building block for the future power grid \cite{Olivares2014}. However, during its operation, a microgrid can experiences disturbances under abrupt events like islanding, load changes, or re-synchronization. Especially, in the islanded mode, voltages and frequency of the microgrid are
no longer supported by a host grid, and thus they must be controlled by DERs such as generators and inverters. With the increasing penetration of inverters in microgrids, it is imperative that the inverters-based DERs should be able to contribute to frequency and voltage regulations, and load sharing in the microgrids under disturbances without causing high circulating var \cite{wei_circulating-current_2017}, an undesired consequence of parallel inverter operation.

To address these challenges, various control strategies have been proposed and have subsequently been aggregated into a hierarchical control architecture \cite{Bidram2012,Olivares2014, trend_2014_taskforce}, which include three levels with different functions: (i) the primary control stabilizes the microgrid and establishes power sharing; (ii) the secondary control removes the deviations in both global frequency and local voltages from the nominal values; (iii) the tertiary control is concerned with global economic dispatch over the network and depends on current energy markets and prices. 

{\color{blue}In primary control, there have been few non-droop based power sharing methods that addresses primary control in a centralized manner and relies on communication to offer precise load sharing as reviewed in \mbox{\cite{trend_2014_taskforce, Bidram2012}}. Some examples of such controls are centralized \mbox{\cite{guerrero2008control, prodanovic2006high}}, master-slave \mbox{\cite{lopes2006defining}}, circular chain control method \mbox{\cite{wu20003c}} etc. 
Nonetheless, in recent standards, the most common and popular primary control is local proportional droop control due to its simple implementation, plug-and-play feature and, reliable performance. It only relies on local measurements and requires no communication as demonstrated in  \mbox{\cite{trend_2014_taskforce, Brabandere_2007_droop, piagi_autonomous_2006, Guerrero_2011_heirarchical}}.}
In secondary control, most of the existing work has focused on centralized approaches due to easier decision making process and coordination \cite{Bidram2012, savaghebi2012secondary, Guerrero_2011_heirarchical}. However, the centralized control architecture does not have the flexibility to integrate the increasing number or variety of new DER devices.
Also, the centralized control architecture requires long communications links, leads to decision delays, and a failure at the control center in extreme events may disrupt the whole system. Therefore, to enable large-scale integration of DERs in power systems, it is necessary to explore decentralized secondary control architecture in microgrids. Recently, some efforts have been pursued on peer-to-peer control architecture for the secondary control \cite{Shafiee2014,Simpson-Porco2015}.
Here, each inverter sends the local measurement signal to its neighboring inverters (i.e., inverters those it has communication with), and then adjusts its dynamics based on the differences between its signal with its neighboring inverters’ signal. Unlike the centralized control, the peer-to-peer control architecture reduces computational resources and allows for plug-and-play integration of DERs (since it does not require the control center to update whenever a new DER device is integrated). 

On the power electronics aspect, the DER inverters are usually of two types, namely grid-forming (GFM) and grid-following (GFL) \cite{Pattabiraman2018,Du2020}. The GFM inverters can directly control voltage and frequency and behave as a controllable voltage source behind a coupling impedance as shown in \figurename \ref{fig:GFL_gfm}(a), which is much like a synchronous generator. Whereas, most of the existing DER inverters are of GFL type that have the capability to control the output current magnitude and phase angle,
and thus, can regulate the real power and reactive power (var) output as shown in \figurename \ref{fig:GFL_gfm}(b). 
{\color{blue}Initially, the common practice was to control GFL inverters only to operate at maximum power point, leaving no headroom for grid services and regulation. However, with improved control schemes and recent grid codes (such as IEEE1547, Rule21, Hawaii’s UL 1741), GFLs are encouraged to operate with certain headroom to provide freq/watt and volt/var regulation \cite{kenyon_stability_2020, hoke_fast_2021, singhal_real-time_2019}. A study \cite{johnson_photovoltaic_2016} demonstrates the economic benefits of implementing freq/watt control in Hawaiian electric utilities by optimal curtailment. Similarly, a recent report \cite{hoke_fast_2021} recommends maintaining GFL headroom either by curtailment or via storage for improved grid reliability in an inverter-dominated system. These developments have taken state-of-the-art to a point where the potential of GFLs is expected to be used in providing grid services beyond the primary controls such as secondary regulation.}


{\color{blue}Two {\bf major gaps in the literature} are identified. The first gap is that GFL inverters, though dominant in practice, are not utilized to provide the secondary control services in a microgrid i.e. voltage and frequency regulation, power sharing and circulating var. The state-of-the-art peer-to-peer secondary control only coordinates GFM inverters for secondary services \cite{Simpson-Porco2015, Shafiee2014}. GFL not being able to contribute to power sharing and frequency/voltage regulation results in under-utilization of their grid support capabilities as well as it may put extra burden on GFM sources in extreme disturbances. The second gap is that almost no study discusses a coordinated control of mixed GFL-GFM sources, at system level, to provide grid services. Most studies have investigated GFM and GFL controllers performance separately only, such as frequency control dynamics in \cite{pattabiraman_comparison_2018, poolla_placement_2019}, fault-ride-through and weak grid synchronization capabilities in \cite{awal_unified_2021} and, device level stability in \cite{fu_large-signal_2021}.}

The objective of the paper is to fill these gaps by proposing GFL-GFM coordination in secondary control and demonstrating its effectiveness by comparing it with uncoordinated and GFM-only-coordinated approaches. We show in this study that if GFLs are not coordinated, their capability remains unutilized, and as a result, system performance is compromised in terms of frequency/voltage regulation as well as power sharing.



In particular, we use a leader-follower consensus (LFC) control architecture to coordinate between GFM and GFL, with the former being leaders and the latter being followers. 
Unlike existing works, we consider both GFM and GFL inveters for frequency/voltage regulations and power sharing. In addition, rather than a standard consensus algorithm, {\color{blue}our work utilizes the LFC algorithm that exploits the physical characteristics of GFM-GFL inverters: (i) the GFM inverters directly control frequency and voltage, and hence, serve as leaders, (ii) the GFL inverters measure frequency/voltage to modulate their outputs, and hence, they serve as followers.} The effectiveness of the proposed LFC coordination is tested on a networked microgrid test system under different disturbances (load changes, intermittency, and, microgrid switching), communication degradation and 'plug-n-play' events i.e., integration or tripping of an inverter. We demonstrate that the proposed fully coordinated control is robust to disturbances, while outperforming uncoordinated and GFM-only-coordinated approaches. Overall, this work emphasizes the necessity and benefits of the GFM-GFL coordination in the secondary control of microgrids.

The structure of this paper is as follows. In Section \ref{sec.model}, we revisit the dynamic models of GFM and GFL inverters and their primary droop controls. Section \ref{sec.control} introduces the LFC framework for frequency/voltage regulation and power sharing among GFM-GFL inverters. Section \ref{sec.validation} presents the co-simulation framework and use cases to compare and demonstrate the effectiveness of the proposed control. Conclusions follow in Section \ref{sec.conclusion}.
\vspace{-0mm}
\section{Modeling of Grid-Forming and Grid-Following Inverter Control Dynamics}
\label{sec.model}
A GFM and GFL inverter can be modeled as an equivalent voltage source and current source, respectively, connected to a distribution network through a filter impedance as shown in \figurename \ref{fig:GFL_gfm}. Both the controls measure terminal voltage ($V^g \angle \delta^g$) and current ($I^g \angle \phi^g$) and regulate the internal voltage $E$ and corresponding angle $\delta$ of an inverter. However, internally, both of these controls have different dynamics as discussed below.
\vspace{-5 mm}
\subsection{Grid-Forming (GFM) Control}
\begin{figure}
    \centering
    \includegraphics[width=1\columnwidth]{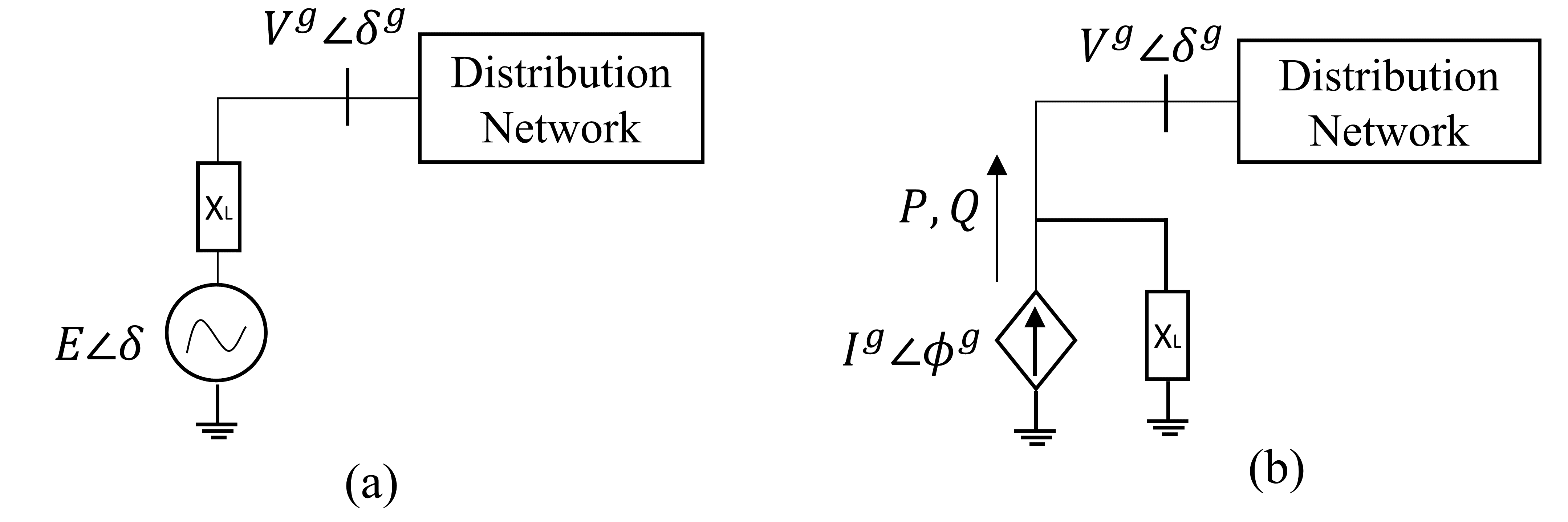}
    \vspace{-8mm}
    \caption{Equivalent representation of inverter controls (a) grid-forming, and (b) grid-following}
    \vspace{-4mm}
    \label{fig:GFL_gfm}
\end{figure}
\begin{figure}
\centering
\includegraphics[width=1\linewidth,trim={0 0 0 0.0cm},clip]{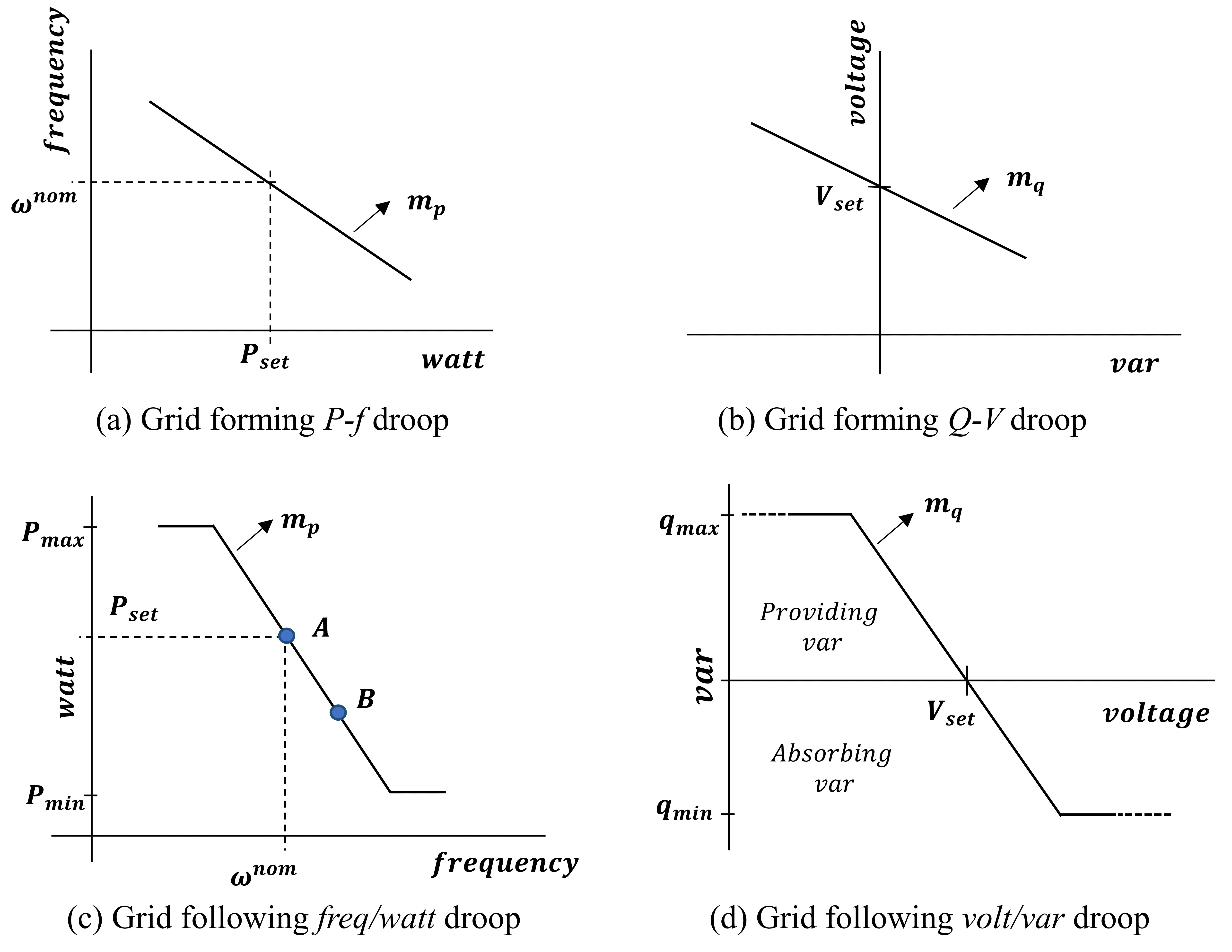}%
\vspace{-3mm}
\caption{Primary frequency and voltage droop control curves of grid-forming and grid-following inverters }
\vspace{-4mm}
\label{fig: droop_curves}
\end{figure}
\begin{figure}
    \centering
    \includegraphics[width=1.05\columnwidth]{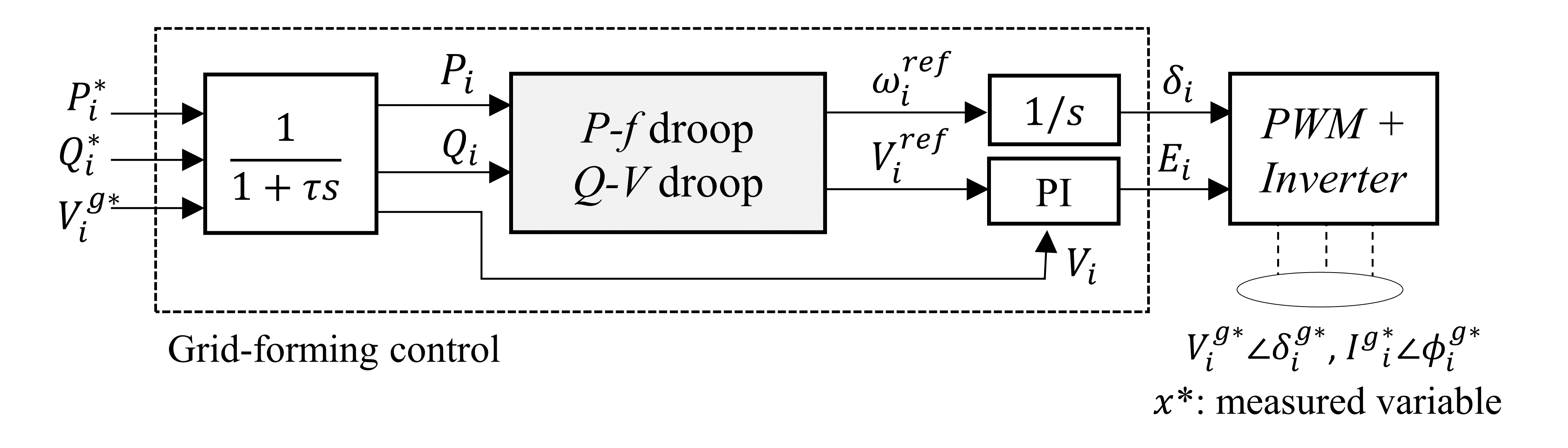}
    \vspace{-9mm}
    \caption{A schematic of grid-forming control dynamics}
    \vspace{-1mm}
    \label{fig:gfm_control}
\end{figure}

An inverter with GFM control can be modeled as an AC voltage source behind a coupling impedance that can generate a desired constant voltage magnitude and frequency in an islanded microgrid as shown in \figurename \ref{fig:GFL_gfm} (a). Mathematically, it can be represented as following for an $i^{th}$ inverter:
\begin{align}
    \dot{\delta_i} &= u_i^{\delta}\\
    E_i &= u_i^{V}
\end{align}
where $u_i^\delta$ and $u_i^V$ are frequency and voltage control inputs or reference signals to the inverter. Several control strategies have been proposed to define the $u_i^\delta$ and $u_i^V$ such as virtual oscillator \cite{johnson_synchronization_2014}, virtual synchronous machine \cite{zhong_virtual_2016} and, droop controls \cite{chandorkar_control_1993,piagi_autonomous_2006}. In this work, we choose a relatively well tested CERTS power-frequency (\textit{P-f}) droop and var-voltage (\textit{Q-V}) droop controls as primary GFM controls \cite{Lasseter2001}. These droop controls regulate the frequency and inverter terminal voltage according to (\ref{eq:pf_droop}) and (\ref{eq:qv_droop}) respectively.
\begin{align}
    \omega_i^{ref} = \omega_i^{nom}-m_{p_i} (P_i-P_i^{set}) \label{eq:pf_droop} \\
    V_i^{ref} = V_i^{set}-m_{q_i} (Q_i-Q_i^{nom}) \label{eq:qv_droop}
\end{align}
where $\omega_i^{ref}$ is a reference frequency and act as control input to the inverter i.e. $u_i^{\delta}=\omega_i^{ref}$. $P_i$ and $Q_i$ denote real and reactive power outputs of inverter respectively. \textit{P-f} control is further defined by 3 parameters i.e. $\omega_i^{nom}$, $m_{p_i}$ and $P_i^{set}$ that represent nominal frequency, droop gain and real power set-point as shown in \figurename \ref{fig: droop_curves}(a). Similarly, in \textit{Q-V} droop curve, $V_i^{ref}$ denotes a reference voltage signal. However, it is not directly used as voltage control input to the inverter ($u_i^{V}$) because we are interested in regulating terminal voltage $V_{i}$ rather than $E_i$ directly. Therefore, $u_i^{V}$ is obtained by passing $V_i^{ref} - V_{i}$ through a PI controller. \textit{Q-V} control is further defined by 3 parameters i.e. $Q_i^{nom}$, $m_{q_i}$ and $V_i^{set}$ that represent nominal var output (usually taken as 0), droop gain and voltage set-point as shown in \figurename \ref{fig: droop_curves}(b). \figurename \ref{fig:gfm_control} shows the full GFM control schematic where measured terminal voltage ($V^{g*}_i$) and output power ($, P_i^*, Q_i^*$) are passed through low pass filter before being used as input to the droop controllers.

 
 The main purpose of \textit{P-f} droop is to enable multiple inverters to share real power. \textit{Q-V} droop helps in mitigating circulating var among paralleled inverters.


\subsection{Grid-Following (GFL) Control}
\begin{figure}
    \centering
    \includegraphics[width=1.0\columnwidth]{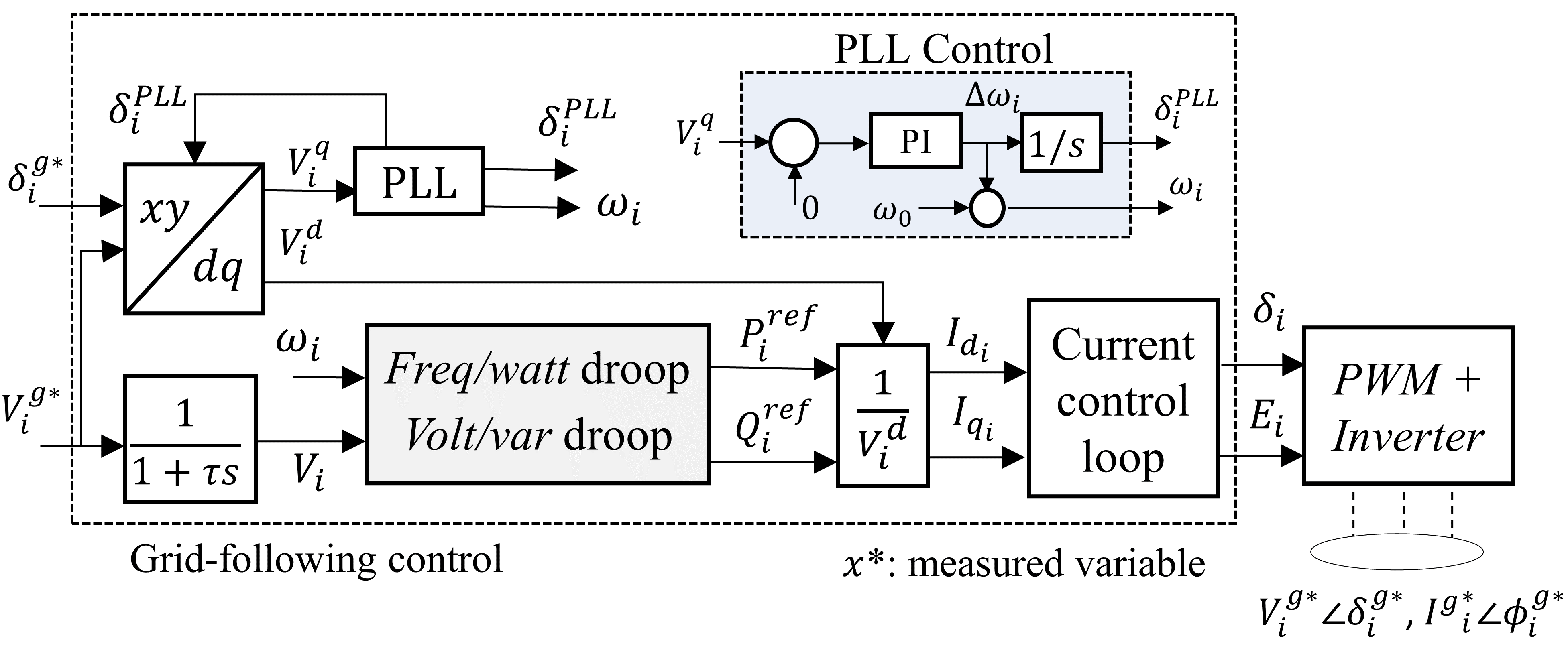}
    \vspace{-7mm}
    {\caption{\textcolor{blue}{A schematic of grid-following control dynamics with droop curves as grid support functions}}}
    \label{fig:gfl_control}
    \vspace{-2mm}
\end{figure}
GFL control enables an inverter to behave like a current source that can inject a specified real power ($P_i^{ref}$) and var ($Q_i^{ref}$) into the grid as shown. In GFL control, it is achieved using a typical current control loop in d-q reference frame which can be written as,
\begin{align}
    I_{d_i} &= P_i^{ref}/V^d_i\\
    I_{q_i} &= Q_i^{ref}/V^d_i
\end{align}
where $I_{d_i}$ and $I_{q_i}$ are reference signals for inverter current in \textit{d-q} reference frame. $V^{d}_i$ is d-axis voltage obtained via \textit{d-q} reference frame transformation that requires both $V_i$ and $\delta^g_i$ as inputs. {\color{blue}Therefore, a standard phase locked loop (PLL) is used to estimate $\delta^g_i$ as $\delta^{PLL}_i$ by constantly regulating q-axis voltage, $V^{q}_i$, to zero using a PI controller as shown in PLL control block in \mbox{\figurename \ref{fig:gfl_control}}} .
PLL is an important component of GFL control to measure grid side frequency i.e. $\omega_i = \dot{\delta}^{PLL}_i$.
Similarly, voltage measurements $V_i^g$ is processed through low pass filter to get $V_i$. Thus obtained current reference signals are converted to inverter source voltage ($E_i, \delta_i$) by the current control loop such that the desired $P_i^{ref}$ and $Q_i^{ref}$ can be injected. The whole schematic sequence of GFL control is shown in \figurename \ref{fig:gfl_control}. The detailed description of PLL and current control loop can be obtained from \cite{Du2020}.

Instead of providing fixed $P_i^{ref}$ and $Q_i^{ref}$ set-points, usually an external grid support function is added to GFLs to indirectly regulate voltage and frequency. In this work, we use DER integration standard IEEE1547-based volt/var and freq/watt droop functions as primary control for GFL inverters as defined below.
\begin{align}
    P_i^{ref} &= P_i^{set} + (\omega^{nom}-\omega_i)/m_{p_i} \label{eq:fr/watt}\\
    Q_i^{ref} &=Q_i^{nom} + (V_i^{set}-V_i)/m_{q_i} \label{eq:volt/var}
\end{align}
Similar to GFM droop curves, $P_i^{set}, \omega^{nom}, m_{p_i}$ and $Q_i^{nom}, V_i^{set}, m_{q_i}$ are control parameters of freq/watt and volt/var droop curves as shown in \figurename \ref{fig: droop_curves}(c) and (d) respectively. It can be observed here that GFM measures power outputs and dispatches frequency and voltage reference signals whereas it gets revered in GFL.

\subsection{Equal Power Sharing}
Equal power sharing is an important controller objective for both GFM and GFL. For a given system condition, an equal power sharing is achieved when all inverters are maintaining the real power dispatch proportional to their droop in pu with respect to their ratings ($s_i$). This can be written as $\eta_p = m_{p_i} P_i / s_i$ for each $i^{th}$ inverter, where $\eta_p$ denotes an equal power sharing parameter. Similarly, an equal var sharing parameter, $\eta_q$ can also be defined. Consequently, an analytical expression for $\eta_p$ and $\eta_q$ can be obtained as,
\begin{align}
    \eta_p = \frac{\sum_{i} P_i}{\sum (s_i/m_{p_i})}; \hspace{5mm} \eta_q = \frac{\sum_{i} Q_i}{\sum (s_i/m_{q_i})}
\end{align}
Any divergence of $P_i$ from $\eta_p$ can be seen as a deviation from equal power sharing. In the case of $Q_i$, a large deviation from $\eta_q$ can cause high circulating vars in the system.

In both GFL and GFM, the main purpose of primary droop controls  
is to maintain the power sharing and prevent circulating var while providing the frequency and voltage regulation. However, due to their proportional nature, the accurate regulation can not be achieved as they always end up with a steady state deviation in both frequency and voltage. A secondary control is needed to accomplish removal of these errors.


\subsection{Need for Coordination among GFL and GFM}
\label{sec:need}
A secondary control is usually applied at GFMs by updating the $P_i^{set}$ in (\ref{eq:pf_droop}). However, in that process, the equal power sharing aspect of primary control gets disturbed. To preserve that, coordination among GFM inverters is needed. 

Further, if GFLs are not coordinated with GFMs' secondary control and left with their primary control only, it will lead to undesired performance as one of the following two situations arises: (a) if GFMs secondary control could restore the frequency to nominal on their own, the GFLs will not see the need of participating in power sharing (point A in \figurename \ref{fig: droop_curves}(c)); thus freq/watt capability of GFLs remains unutilized or; (b) if GFMs do not have enough capacity to manage the disturbance on their own, GFLs do participate, but it prevents frequency restoration (point B in \figurename \ref{fig: droop_curves}(c)). Moreover, in both situations, power sharing also gets disturbed as explained in the case study later. This necessitates a control approach that coordinates among GFL and GFM, which is explored in the next section. 

\vspace{-1mm}
\section{Leader-Follower Consensus (LFC) Architecture for GFL-GFM Coordination}
\label{sec.control}

A special consensus framework, namely LFC, will be utilized in this section to implement a peer-to-peer coordinated secondary control for a fleet of GFL and GFM inverters. In the standard consensus framework, multiple agents aim to reach an agreement on some decision by exchanging information among their peer agents. Then each agent adjusts its dynamics or behavior based on the information received from the peer agents with whom it communicated with \cite{Olfati-Saber2006}. In the LFC framework, one or more agents are selected as leaders with external inputs in addition to the  consensus protocol, while the remaining agents are followers only obeying the  consensus protocol \cite{Jadbabaie2003}. 

Mathematically, a group of agents can be denoted as  $A = \{1,2,...,n\}$, in which the first $n_f$ agents are selected as followers while the last $n_l$ agents are selected as leaders with respective notations  $A_F = \{1,2,...,n_f\}$, $A_L = \{n_f + 1,n_f + 2,...,n_f +n_l\}$ and $n=n_f +n_l$. Consider $\mathcal{N}_i$ as the set of agents that communicate with agent $i$. For simplicity, let $x_i \in R$ be the position of agent $i$, where we only consider the one dimensional case, yet the framework can be extended to higher dimensions. The state evolution of each follower $i \in A_F$ is governed by the first order consensus protocol:
\begin{align}
\dot{x}_i = \sum_{j\in \mathcal{N}_i} (x_j-x_i), i \in A_F,  
\end{align}
while the state evolution of each leader $i \in A_L$ is governed by the first order agreement protocol with an assigned external input $u_i \in R$:
\begin{align}
\dot{x}_i = \sum_{j\in \mathcal{N}_i} (x_j-x_i)  + u_i, i \in A_L.
\end{align}
The external control input $u_i$ can be designed to achieve additional control objectives, e.g., ensuring the agents reach the agreement in one predefined way \cite{Macellari2017}.

{\color{blue}Since GFM inverter control actively sets the frequency and voltage whereas GFL control measures them to modulate its output, their combination naturally fits the leader-follower architecture, in which GFM inverters serve as leaders and GFL inverters serve as followers.} 
Consider a microgrid network with $N_L$ GFL and $N_M$ GFM inverters represented by sets $L=\{1,2,3, \dots , N_L\}$ and $M=\{N_L+1, N_L+2, N_L+3,\dots, N_L+N_M\}$ respectively, with total number of inverters denoted by $N=N_L+N_M$. The communication link pattern among the inverters can be represented by an $N \times N$ \textit{adjacency matrix} $C$, with elements $c_{ij}$, i.e., $c_{ij} = 1$ if the communication exists between inverter $i^{th}$ and $j^{th}$, and $c_{ij} = 0$ otherwise. {\color{blue}It can be referred from graph theory that for a group of agents to converge to a common consensus, the communication network among them must be \textit{connected} i.e. there must be a path in the communication graph between any two nodes  \cite{ren_information_2007, olfati-saber_consensus_2007}. In a system of $N$ agents, the minimum and maximum communication links required to keep graph connected are $N-1$ and $N\times(N-1)/2$, respectively.}
In the next two subsection, we apply the leader-follower concept to design a coordinated secondary controller for GFL and GFM inverters.
\vspace{-3.5mm}
\subsection{Frequency Regulation and Real Power Sharing}
There are two objectives to achieve while designing secondary control for \textit{P-f} (GFM) and freq-watt (GFL) droop controls i.e. frequency restoration to nominal and equal real power sharing among inverters. To accomplish this, we propose following secondary control for GFL (\ref{eq:freq_sec_GFL}) and GFM (\ref{eq:freq_sec_gfm}):
\begin{align}
     -k_i^p \frac{dP_i^{set}}{dt} &= \sum_{j\in \mathcal{N}_i}{c_{ij}(m^p_iP_i^{set} - m^p_jP_j^{set})} , \forall \; i\in L \label{eq:freq_sec_GFL}\\
    -k_i^p \frac{dP_i^{set}}{dt} &= (\omega_i - \omega^{nom}) + \!\!\sum_{j\in \mathcal{N}_i}{c_{ij}(m^p_iP_i^{set} - m^p_jP_j^{set})}, \label{eq:freq_sec_gfm}
    \\ &\qquad\qquad \forall \; i\in M  \nonumber
\end{align}
Here, $k_i^p$ is a positive gain for $i^{th}$ inverter and affects the speed of frequency regulation. $P_i^{set}$ from primary controls is used as secondary control variable. Following leader-follower concept, both GFL and GFM have real power sharing terms whereas only GFM has an extra frequency restoration term in (\ref{eq:freq_sec_gfm}). Thus, (\ref{eq:pf_droop})+(\ref{eq:freq_sec_gfm}) and (\ref{eq:fr/watt})+(\ref{eq:freq_sec_GFL}) make a complete frequency control for GFM and GFL respectively. Note that with no communication, this control will be an equivalent of frequency restoration without ensuring power sharing what we term as uncoordinated secondary control.

{\color{blue}\textit{Remark 1}: 
Our work assumes that GFL sources will have a planned headroom, required for secondary regulation, based on the recent standards and recommendations (discussed in Section I) that encourage non-dispatchable GFL inverters (such as PV) to participate in frequency regulation by maintaining certain headroom either via storage or curtailment \cite{hoke_fast_2021, johnson_photovoltaic_2016}. The optimal headroom planning is an important aspect, though not the focus of this work, and can be referred from some recent proposed methods \cite{yuan_machine_2020,hoke_rapid_2017}.}

{\color{blue}\textit{Remark 2}:
Even in the case of planned headroom, intermittency in the GFL source may cause an unexpected reduction in the available headroom. In that situation, the proposed control will be able to converge and achieve full frequency restoration as long as other inverters together have sufficient headroom to compensate for the intermittency, though with compromised power sharing. Let's assume for $i^{th}$ GFL inverter, the output $P_i$ is saturated to a lower value $P_i^{max}$ while the power required for equal power sharing is $P_i^{eq}$ such that $P_i^{max}<P_i^{eq}$. In this case, equal power sharing is not possible. However, that information is not known to the GFL controller (\ref{eq:freq_sec_GFL}), and it still converges assuming the original droop set-point $P_i^{set}=P_i^{eq}$. It helps the controller to converge, whereas at the device level, the actual output $P_i$ is saturated to $P_i^{max}$ and the remaining power is compensated by the other non-saturated inverters. }

\vspace{-2mm}
\subsection{Voltage regulation and reactive power sharing}
Similar to previous section, we have two objectives here i.e. voltage regulation and mitigating circulating var by equal var sharing. However, unlike frequency regulation and real power sharing, it is not possible in general to achieve both the objectives accurately. We can achieve the desired internal voltage $E_i$ but terminal voltage $V_i$ is impacted by network impedance. Therefore, an equal var sharing is not always possible with the accurate voltage regulation in parallel inverter operations as observed and explained by \cite{Simpson-Porco2015, tuladhar_parallel_1997}. Therefore, we design a compromise between two objectives as following:
\begin{align}
    -k_i^q \frac{dV_i^{set}}{dt} &= \sum_{j\in \mathcal{N}_i}{c_{ij}(m^q_iQ_i - m^q_jQ_j)}, \forall \; i\in L \label{eq:volt_sec_GFL} \\
    -k_i^q \frac{dV_i^{set}}{dt} &= \alpha (V_i\! -\! V^{nom}) +\! \beta \! \sum_{j\in \mathcal{N}_i}{\!\!c_{ij}(m^q_iQ_i - m^q_jQ_j)}, \nonumber
    \\ & \qquad\qquad \forall \; i\in M  \label{eq:volt_sec_gfm}
\end{align}
Here, $k_i^q$ is a positive gain and $V_i^{set}$ is a secondary control variable from primary controls. $V^{nom}$ is a nominal voltage usually taken as 1 pu. The trade-off between voltage regulation and var sharing is implemented by scaling coefficients $\alpha$ and $\beta$ for GFM in (\ref{eq:volt_sec_gfm}). The tuning of the scaling coefficient will depend on the network and utility requirements. The tuning can also be made adaptive, though we use offline tuning in this study. Thus, (\ref{eq:qv_droop})+(\ref{eq:volt_sec_gfm}) and (\ref{eq:volt/var})+(\ref{eq:volt_sec_GFL}) make a complete voltage control for GFM and GFL respectively. Note that with no communication, it will be an equivalent of accurate voltage regulation with high circulating var.

\vspace{-2mm}

\section{Co-simulation for Validation}
\label{sec.validation}
\vspace{-0.5mm}

\begin{figure}
    \centering
    \includegraphics[width=1.05\columnwidth]{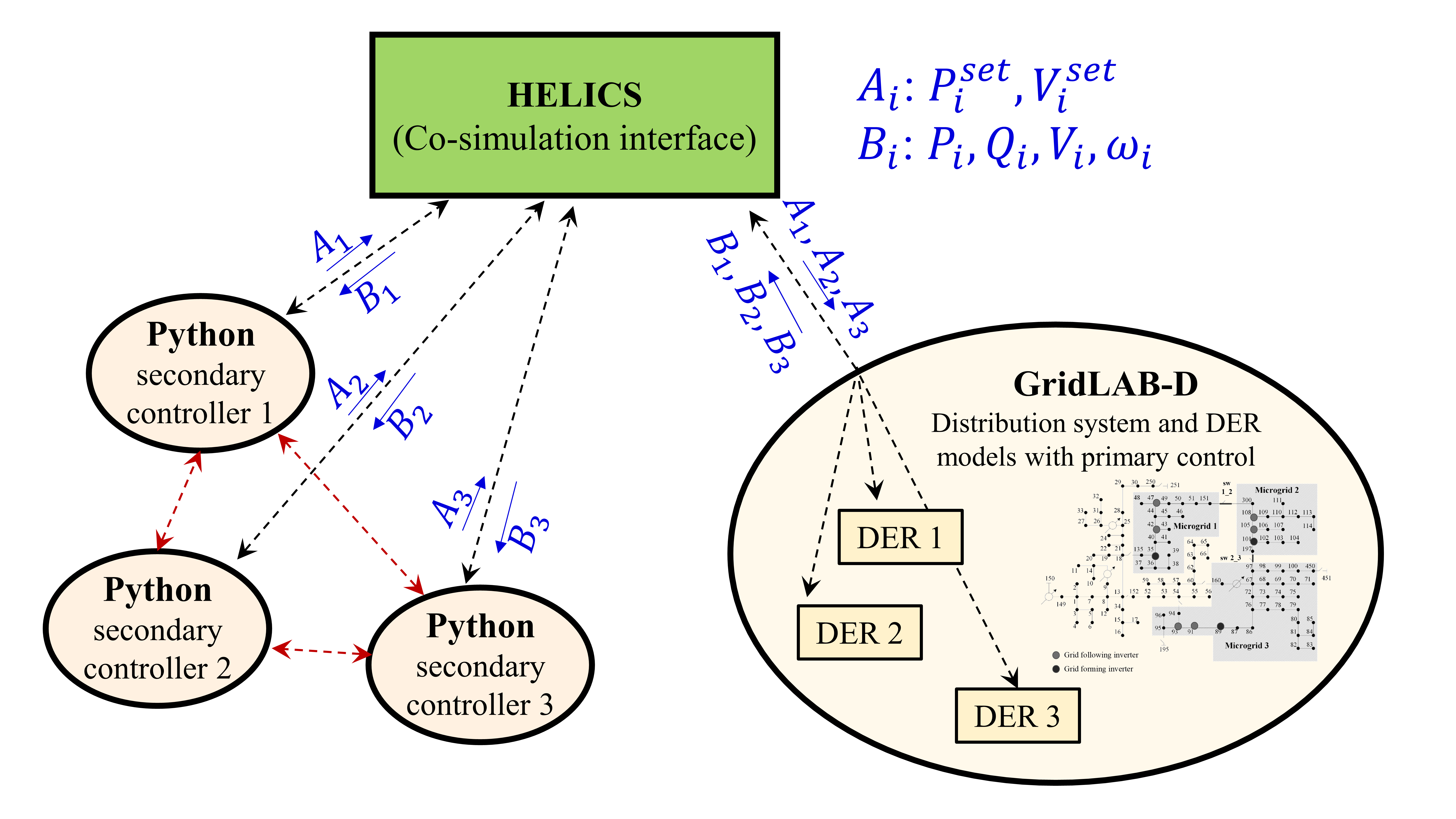}
    \vspace{-2mm}
	\caption{Co-simulation platform for implementing DER coordination control}
	\vspace{-2mm}
	 \label{fig.co-sim}
\end{figure}
\begin{figure}
    \centering
    \includegraphics[width=0.95\columnwidth]{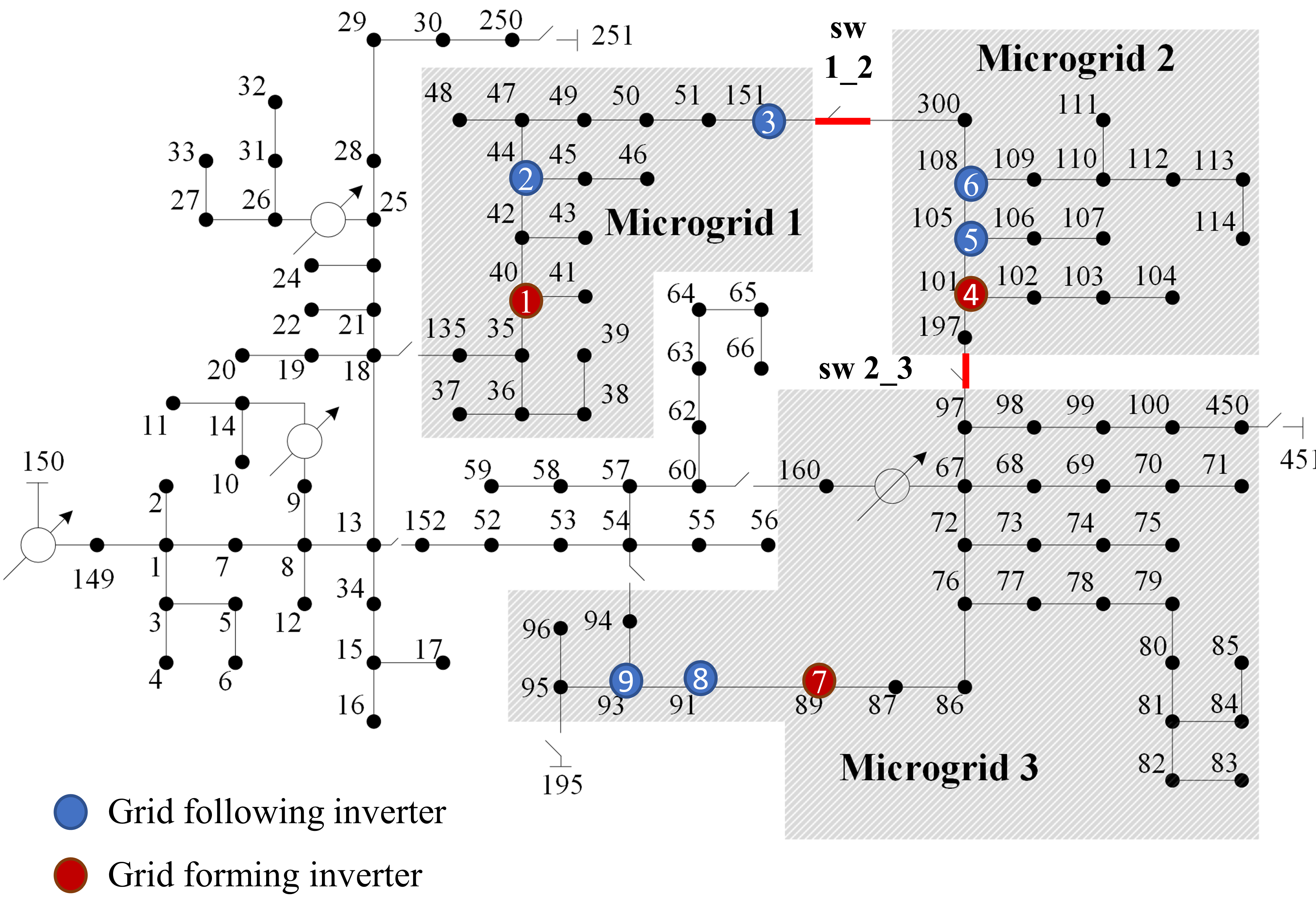}
    \vspace{-3mm}
    \caption{A modified IEEE 123 test feeder to create a networked microgrid system}
    \label{fig:test_system}
    \vspace{-5mm}
\end{figure}

We develop a co-simulation platform as presented in Fig. \ref{fig.co-sim}, in which the secondary peer-to-peer controllers are programmed in Python to send control action to the respective DER inverters. The microgrid modeling and inverter dynamics are simulated in GridLAB-D \cite{chassin_gridlab-d_2014}. An open-source middleware HELICS \cite{palmintier_design_2017} is used to handle the data exchange between GridLAB-D and the Python-based controllers, and to maintain time synchronization between the individual programs.

A modified and fully inverter-based IEEE 123-node test feeder is used to demonstrate the impact of the proposed LFC control. As shown in \figurename \ref{fig:test_system}, it is a networked microgrid test system in which three microgrids (shown in the shaded area) are interconnected via switches $sw$ ${1\_ 2}$ and $sw$ ${2 \_ 3}$. There are total 9 utility-scale inverters installed in the system, out of which 6 are GFL, and 3 are GFM inverters at nodes shown in \figurename \ref{fig:test_system}. Each GFM and GFL has ratings of 600 and 350 kW respectively. Note that each microgrid has 1 GFM and 2 GFL inverters. The total rating of the inverters is about 3900 kW, and the total peak load in the networked microgrid is about 3500 kW. All inverters have 1\% frequency droop and 5\% voltage droop values. In order to test the control performance in a very low-inertial microgrid, there is no generator installed in the system.

To demonstrate the need for coordination among GFL and GFM, the performance is compared with three other control strategies as following: (a) Case I (\textit{no-control}), where only primary droop control is deployed with no secondary control; (b) Case II (\textit{un-coordinated}), in which local secondary control is deployed with no communication among inverters; (c) Case III (\textit{GFM-coordinated}), in which only GFM inverters are coordinated as proposed in \cite{Simpson-Porco2015} but GFL inverters are left uncoordinated; (d) Case IV (\textit{fully coordinated}) in which GFL and GFM inverters are coordinated with proposed LFC control.

In order to evaluate power sharing performance, we define the mean power sharing index for real power (MPSI) and var (MQSI) as following:
\vspace{-4mm}
\begin{align}
    MPSI = \sum_{i=1}^{N}|(m_{p_i} P_i / s_i - \eta_p)/\eta_p|/N\\
    MQSI = \sum_{i=1}^{N}|(m_{q_i} Q_i / s_i - \eta_q)/\eta_Q|/N
\end{align}
\vspace{-1mm}
Lower values (close to 0) of MPSI and MQSI are desired as they represent an average deviations of inverters dispatch from equal power. A higher MQSI is a measure of high circulating var in the system.
\vspace{-3.5mm}
\begin{figure}
    \centering
    \includegraphics[width=0.85\columnwidth]{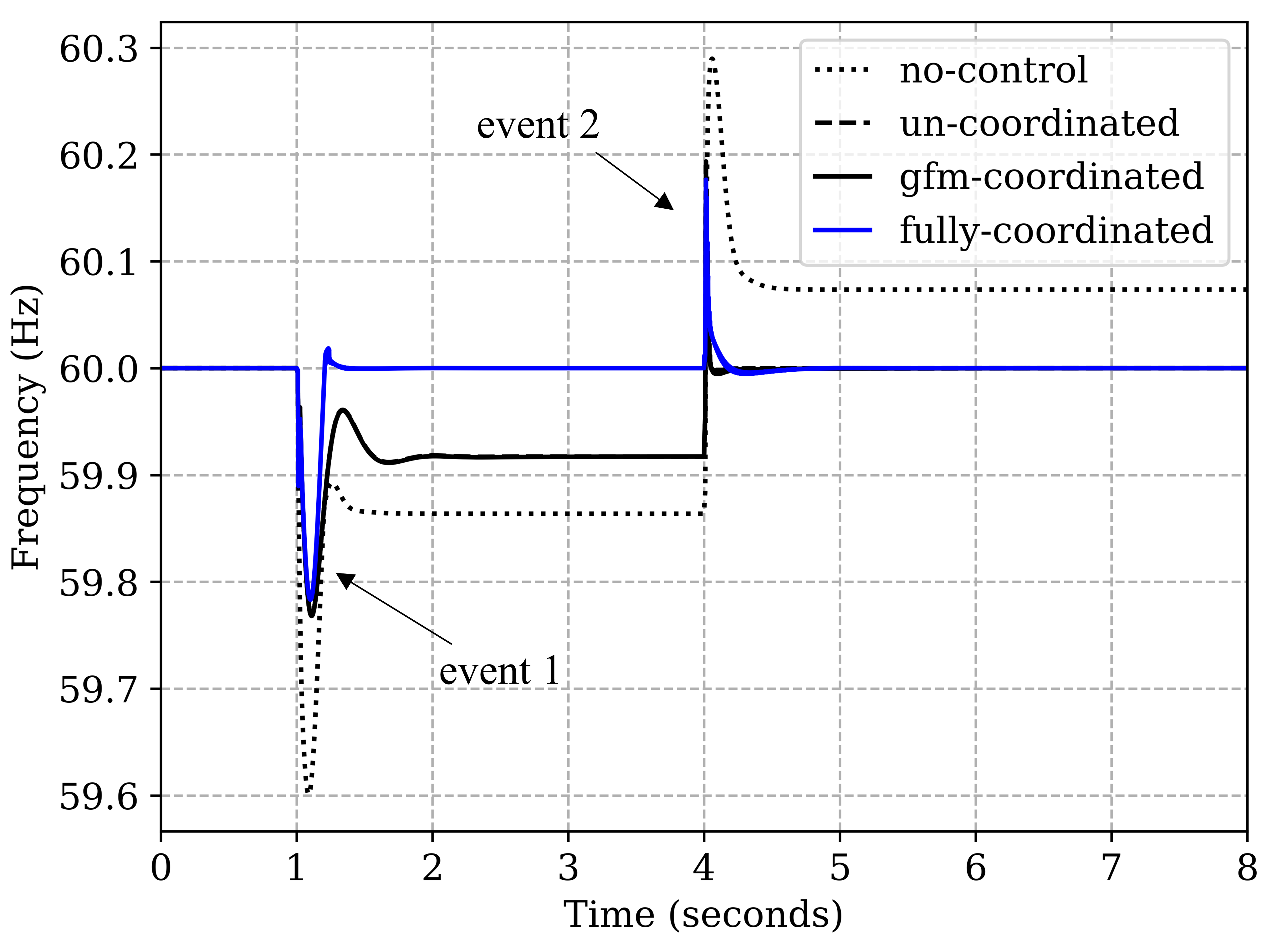}
    \vspace{-2mm}
	\caption{Comparison of the frequency response by proposed LFC control with existing secondary frequency controls in the events of islanding and load disturbance. LFC outperforms all other secondary controls in the event 1.}
	\vspace{-2mm}
	 \label{fig:freq_compare}
\end{figure}

\begin{figure}
    \centering
    \includegraphics[width=0.9\columnwidth]{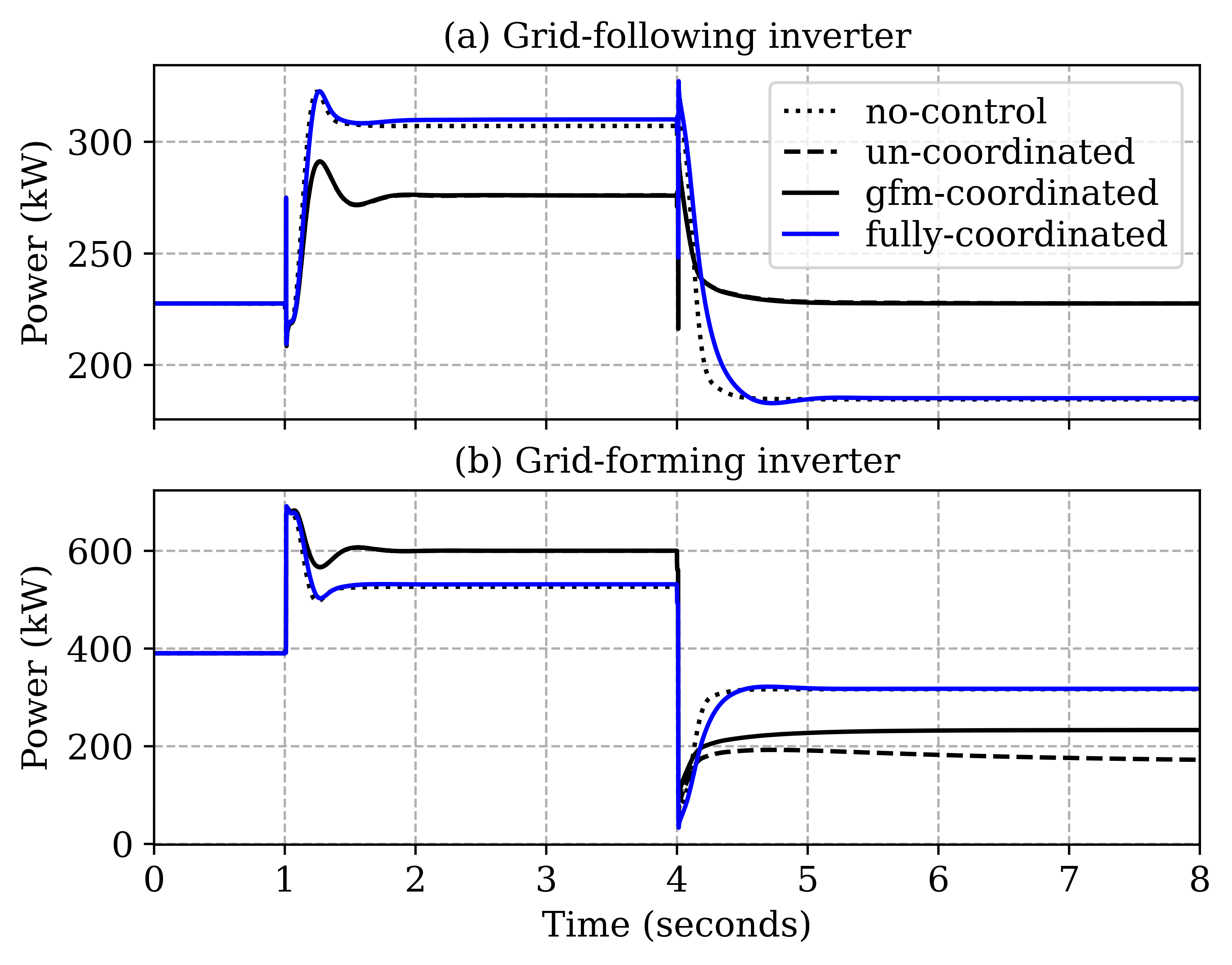}
    \vspace{-3mm}
	\caption{Real power response of GFL and GFM inverters to frequency deviation }
	\vspace{-2mm}
	 \label{fig:p_GFL_compare}
\end{figure}

\begin{figure}
    \centering
    \includegraphics[width=0.95\columnwidth]{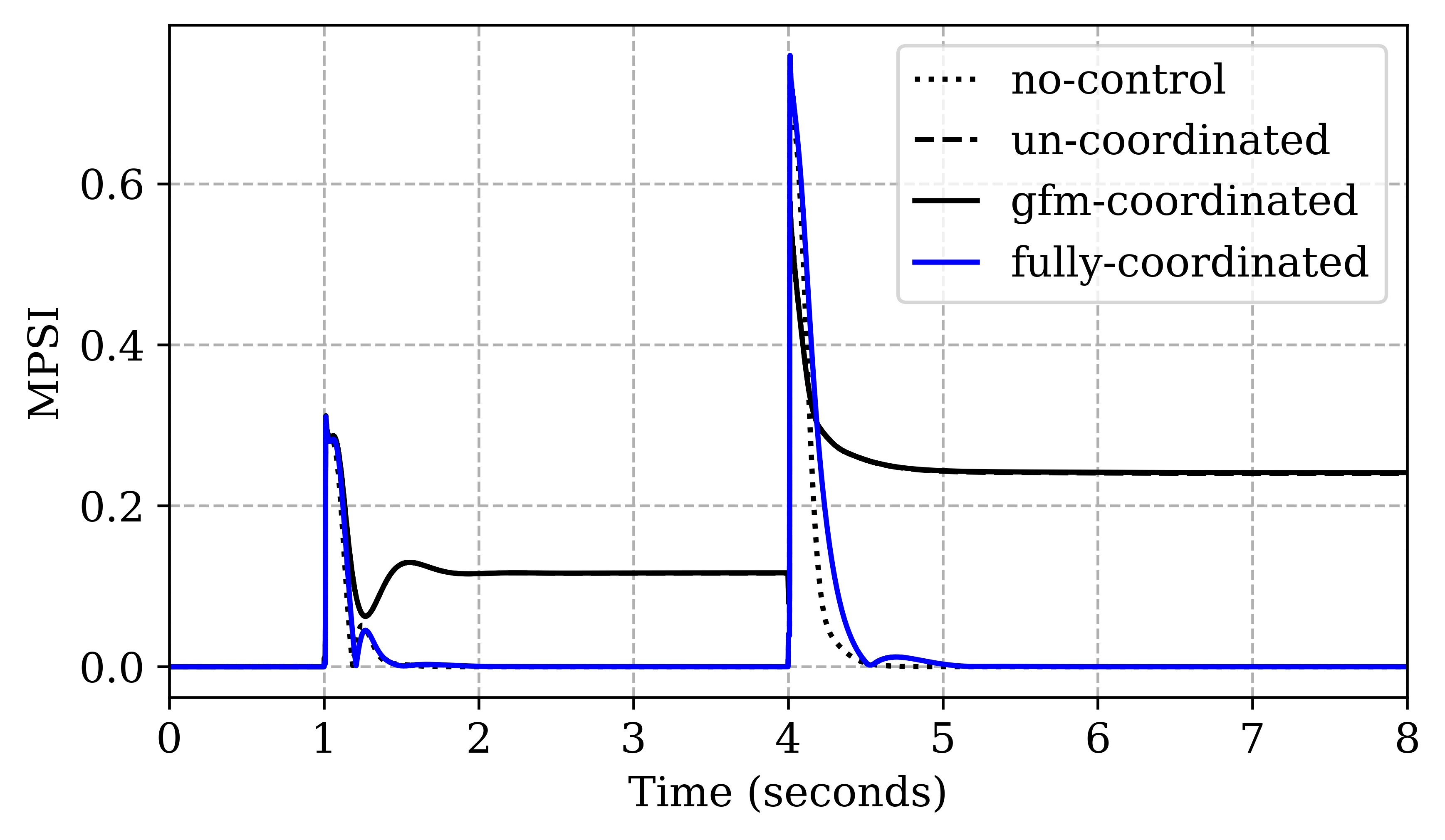}
    \vspace{-4mm}
	\caption{Real power sharing index performance comparison of LFC control with existing secondary frequency controls. LFC significantly outperforms both un-coordinated and GFM-coordinated secondary controls in both events.}
	\vspace{-4mm}
	 \label{fig:mpsi_compare}
\end{figure}

\subsection{Frequency Regulation and Real Power Sharing}
We create two disturbance events (at $t=1$ and $t=4$) to emphasize the value of the full coordination of GFL-GFM inverters. To begin with, at $t=0$, the whole system is connected to the grid via substation. At $t=1$ (event 1), the networked microgrid (shaded area) is isolated from the rest of the system and substation by opening switch between bus 13 and 152. At $t=4$ (event 2), a load disturbance is simulated by opening switch between bus 60 and 160 that disconnects around 1200 kW load (35\%). The frequency responses to these events in all 4 cases are compared in \figurename \ref{fig:freq_compare}. The islanding event at $t=1$ leads to a dip in frequency followed by only partial recovery to 59.86 Hz by primary droop control in Case I. It can be seen that only the proposed LFC control (Case IV, blue line) is able to recover the frequency fully back to 60 Hz. The Case II and III control strategies could only restore frequency to 59.92 Hz. However, in response to event 2 at $t=4$, Case II and III are able to recover the frequency overshoot back to 60 Hz along. The different responses to these two events in Case II and III can be attributed to the non-coordination of GFLs as discussed in Section \ref{sec:need}. The event 1 disturbance was large enough to require both GFL and GFM inverters to share their power to meet the total load demand of the microgrid. Therefore, GFLs participate but prevent frequency restoration. Whereas, in event 2, since GFM inverters were enough to manage the load disturbance and could recover the frequency to 60 Hz, GFL don't see the need for participation and settle at their nominal power without sharing the load as can be observed in \figurename \ref{fig:p_GFL_compare}(a). 
 
Consequently, the power sharing performance (MPSI) can be seen in \figurename \ref{fig:mpsi_compare}. LFC control is able to achieve 0 MPSI i.e. equal power sharing in less than 1 second in both the events whereas Cases II and III have non zero MPSI in both disturbances. Note that the non-participation of GFL in event 2 for case II and III leads to a worse power sharing reflected by a higher MPSI value, compared to event 1. Table \ref{tab:freq_power_index} summarizes the frequency regulation and real power sharing performance comparison among all cases. The coordinated LFC control (Case IV) is able to achieve frequency restoration as well as accurate power sharing in both events. 

 This use case demonstrates the need for coordination and verifies that if GFL and GFM left uncoordinated (Case II and III), the inverters capabilities remain under-utilized and controller performance is compromised. 
 
 \begin{table}
\renewcommand{\arraystretch}{1.15}
\caption{Frequency error and real power sharing index comparison }
\label{tab:freq_power_index}
\centering
\begin{tabular}{c c c  c c}
\hline
\multirow{2}{*}{Case Type} & \multicolumn{2}{c}{Event 1} & \multicolumn{2}{c}{Event 2}\\
  & $|f-60|$  & $MPSI$ & $|f-60|$  & $MPSI$ \\
\hline
no-control & 0.14 & 0.00 & 0.07 & 0.00\\
un-coordinated & 0.08 & 0.12 & 0.00 & 0.24\\
GFM-coordinated & 0.08 & 0.12 & 0.00 & 0.24\\
fully coordinated & 0.00 & 0.00 & 0.00 & 0.00\\
\hline
\end{tabular}
\vspace{-2mm}
\end{table}

\begin{figure}
    \centering
    \includegraphics[width=0.85\columnwidth]{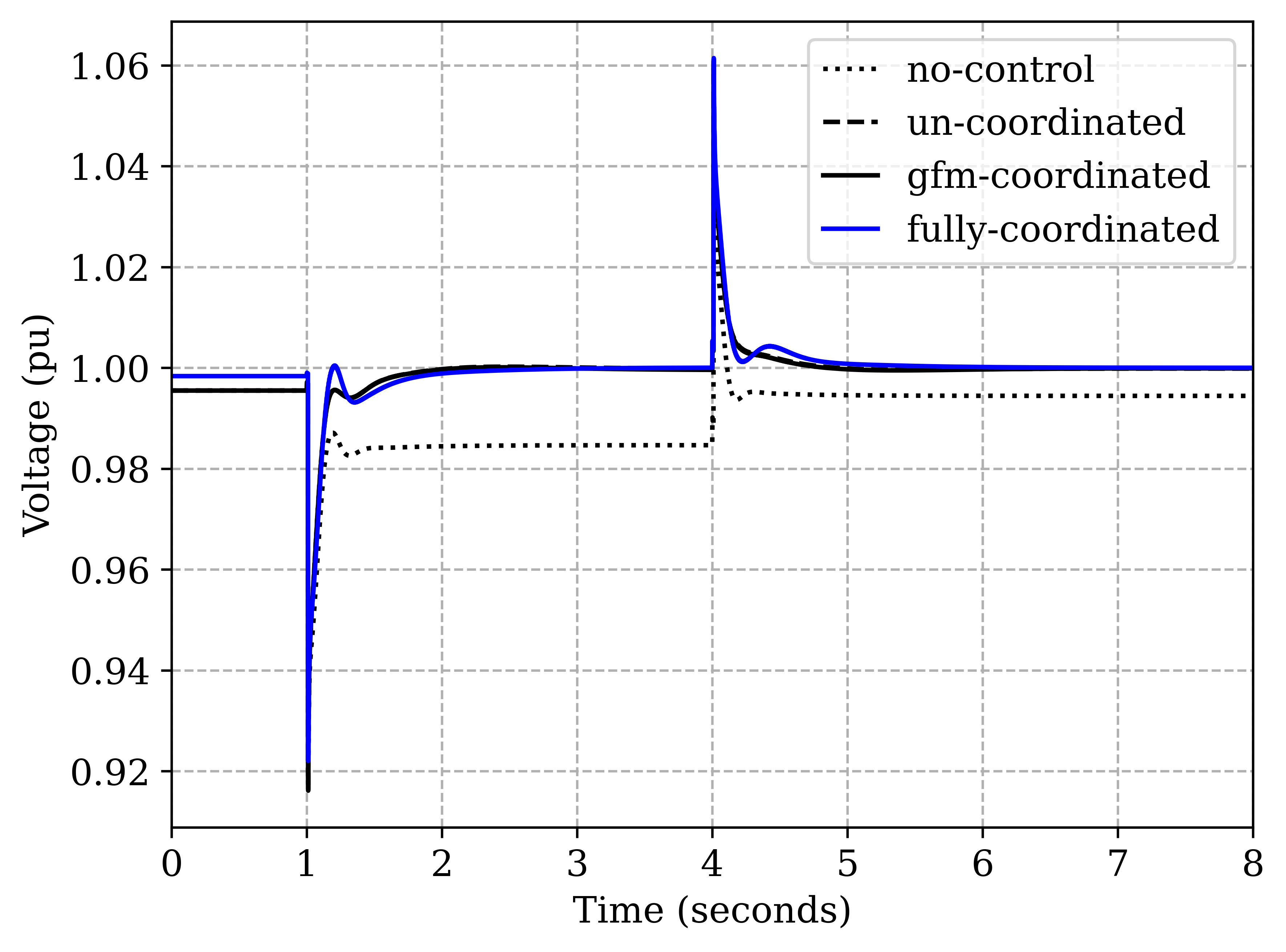}
    \vspace{-2mm}
	\caption{Comparison of the voltage response by proposed LFC control with other secondary voltage controls in the events of islanding and load disturbance.}
	\vspace{-0mm}
	 \label{fig:volt_compare}
\end{figure}


\begin{figure}
    \centering
    \vspace{-2mm}
    \includegraphics[width=0.9\columnwidth, trim={0 0 0 5}, clip]{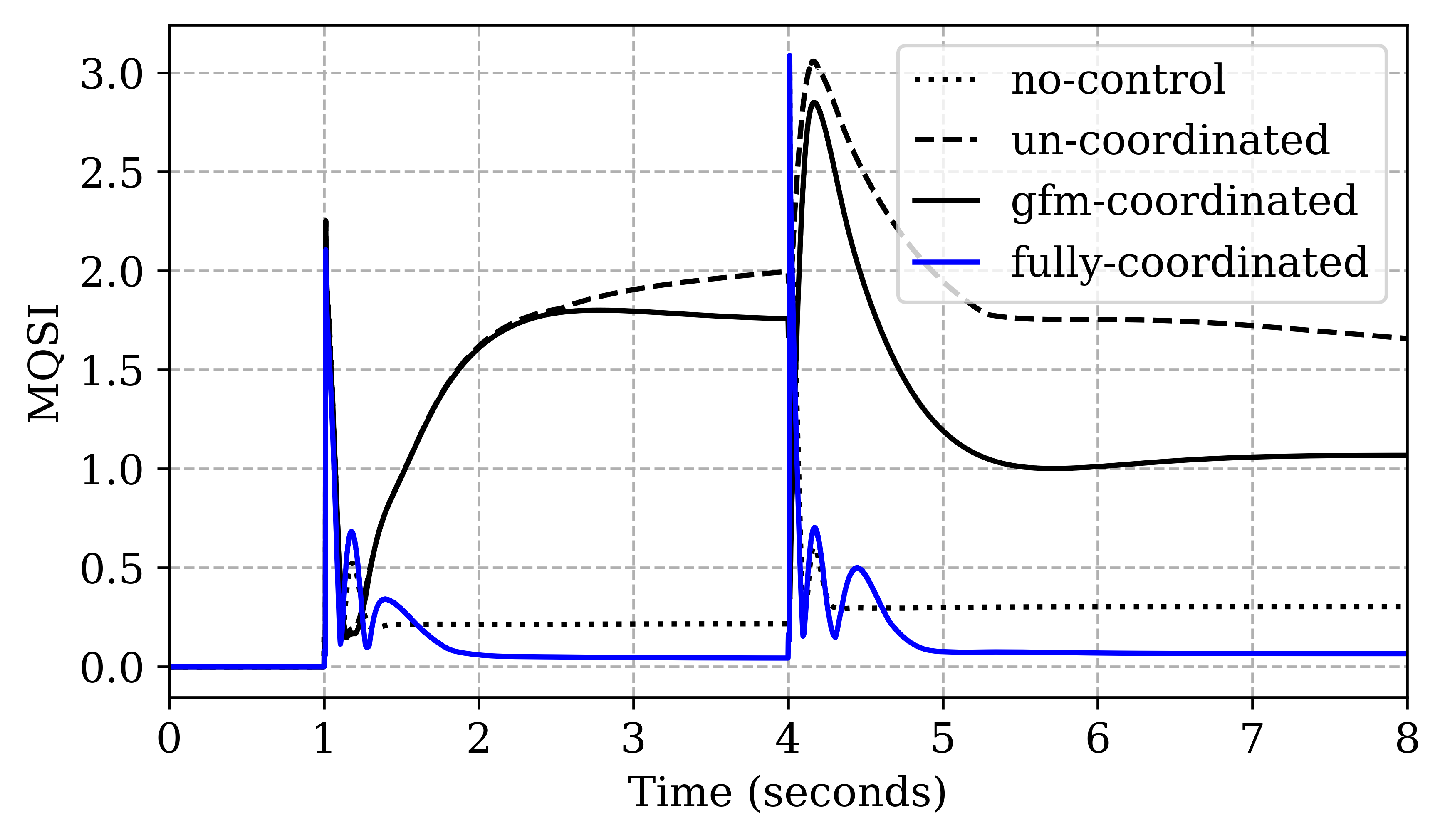}
    \vspace{-4mm}
	\caption{Var sharing index performance comparison of LFC control with other secondary control approaches. Clearly, the LFC with full-coordination of GFM-GFL inverters significantly outperforms other secondary control approaches equal var sharing among inverters.}
	\vspace{-2mm}
	 \label{fig:mqsi_compare}
\end{figure}

\subsection{Voltage Regulation and Var sharing}
\vspace{-0mm}
For the same disturbance events, an average voltage responses and var sharing performance (MQSI) in all 4 cases are compared in \figurename \ref{fig:volt_compare} and \figurename \ref{fig:mqsi_compare}, respectively.  
On investigating the first event at $t=1$, we observe that primary droop controls are not able to bring voltages to nominal 1 pu in case I. Further we observe in \figurename \ref{fig:volt_compare} that Case II and III are able to improve voltage regulation significantly due to GFM's secondary volt/var control. However, it leads to a very high amount of circulating var as reflected from higher MQSI values compared to the case I, shown in \figurename \ref{fig:mqsi_compare}. This undesired response occurs because each inverter tries to push it's own voltage to 1 pu by injecting more var with no coordination with other inverters in case II or with GFL in case III. Note that Case III MQSI value is relatively better than Case II because of partial coordination.

On the other hand, the proposed LFC coordinated control in Case IV arrives at a compromise between voltage regulation and var sharing. The voltage is very close to nominal voltage as shown in \figurename \ref{fig:volt_compare} while maintaining MQSI close to 0 in steady state as shown in \ref{fig:mqsi_compare}. Table \ref{tab:volt_regulation} summarizes the performance trade-off, where the voltage regulation is evaluated by calculating a mean absolute error over all inverters in steady state, defined as $V_{error} = {\sum_{i=1}^{N} |V_{i_t} - 1|}/{N}$, where $N$ denote the total number of inverters. It can be verified that the proposed Case IV is able to improve both $V_{error}$ and MQSI significantly (more than 5 times) compared to Case I. Whereas, in both Case II and III, a highly accurate voltage regulation cost us significantly higher MQSI.

The second event at $t=4$ turns out to be a favorable disturbance for voltages i.e. smaller $V_{error}$ in all cases. Nonetheless, the superior performance of LFC control in voltage regulation as well as var sharing can be observed in this event as well from Table \ref{tab:volt_regulation}.

\begin{table}
\renewcommand{\arraystretch}{1.15}
\caption{Voltage regulation and var sharing index comparison }
\label{tab:volt_regulation}
\centering
\begin{tabular}{c c c  c c}
\hline
\multirow{2}{*}{Case Type} & \multicolumn{2}{c}{Event 1} & \multicolumn{2}{c}{Event 2}\\
  & $V_{error}$  & $MQSI$ & $V_{error}$  & $MQSI$ \\
\hline
no-control & 0.015 & 0.22 & 0.006 & 0.30\\
uncoordinated & 0.001 & 2.03 & 0.000 & 1.66\\
GFM-coordinated & 0.002 & 1.77 & 0.002 & 1.07\\
fully-coordinated & 0.002 & 0.04 & 0.001 & 0.05\\
\hline
\end{tabular}
\vspace{-2mm}
\end{table}

\vspace{-2mm}

\color{blue}
\subsection{Impact of intermittency in GFL}
\begin{figure}
    \centering
    \includegraphics[width=0.95\columnwidth]{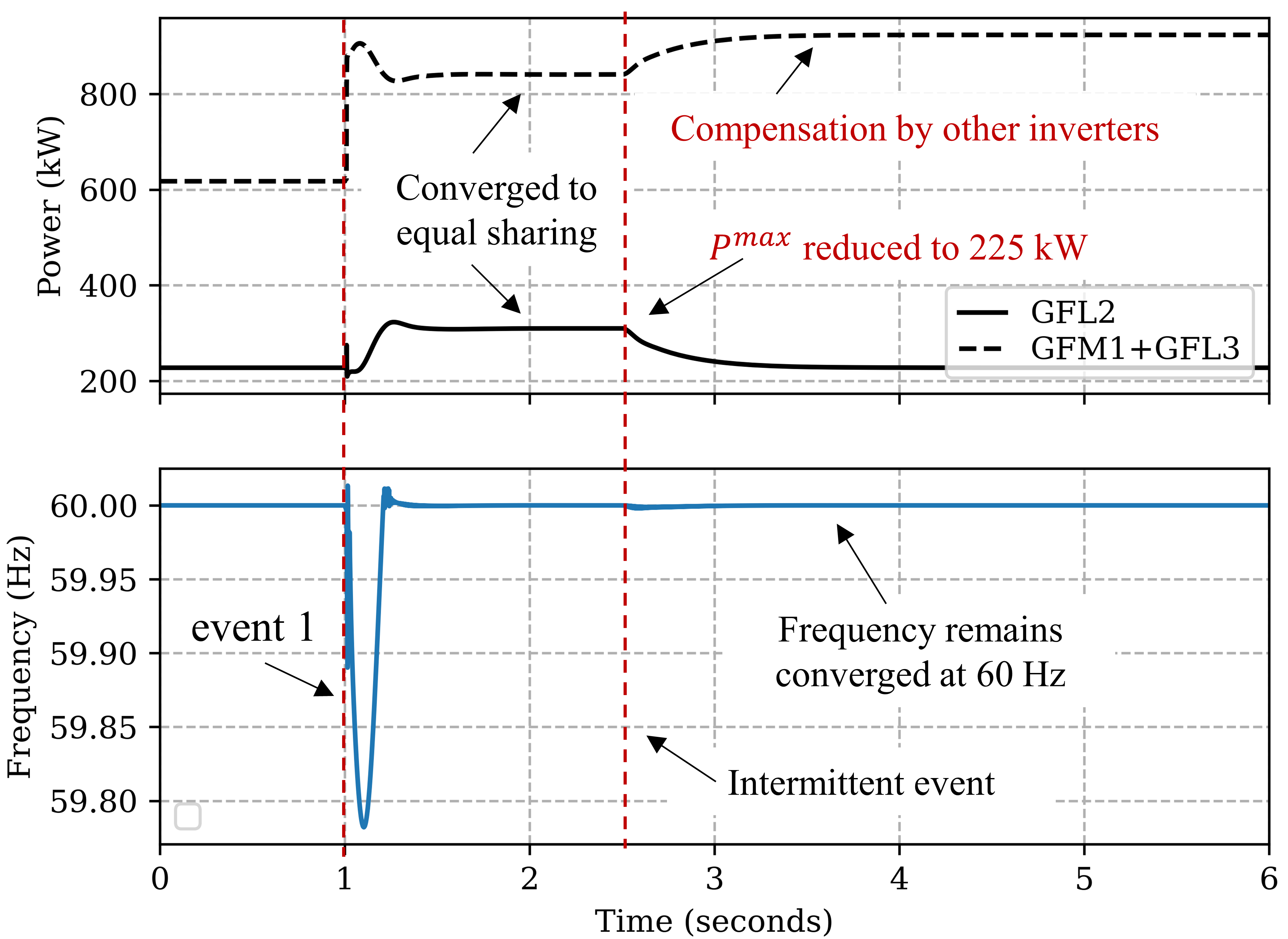}
    \vspace{-1mm}
	\caption{\textcolor{blue}{Impact of an intermittent event on control performance when the GFL inverters 2,5,8 are not fully dispatchable}}
	\vspace{-4mm}
	 \label{fig:int}
\end{figure}

The impact of intermittent nature of GFL sources is observed by simulating an intermittency event (e.g., cloud cover) on GFL 2, 5 and, 8 at t=2.5 that reduces their maximum power output by from 350 kW to 225 kW.  It can be observed in \figurename \ref{fig:int} (top), where after the islanding event 1 at t=1, all inverters converge to an equal power sharing dispatch. At t=2.5, due to intermittency, GFL 2 power converges to 225 kW while inverters (1 and 3) compensate for it. A similar behavior is followed by other inverters as well. Along with convergence, a full restoration of frequency to 60 Hz is also maintained in the intermittent event as shown in \figurename \ref{fig:int} (bottom). However, power sharing is disturbed, reflecting a lower MPSI value of 0.15, as reported in Table \ref{tab:int}. It compares the impact of intermittent events of different magnitude on power sharing and frequency deviation. It can be observed that higher intermittency worsens the power sharing with no impact on frequency deviation. The system can only handle the $P^{max}$ reduction of intermittent GFLs till 200 kW due to limited headroom availability with all other inverters at t=2.5. Further note that, in these intermittencies, with reducing real power headroom, there is more room for reactive power in the inverter. Therefore, the voltage regulation is not affected negatively by intermittency as observed in Table \ref{tab:int}.

\begin{table}
\renewcommand{\arraystretch}{1.1}
\caption{\color{blue}Impact of intermittent events of different magnitude on the frequency deviation and equal power sharing indices}
\label{tab:int}
\centering
\begin{tabular}{c c c c c c}
\hline
 \multirow{2}{*}{$P^{max}$} & No intermittency & \multicolumn{4}{c}{Intermittency (kW)}\\
  & 350 kW (rated) & 275 & 250 & 225 & 200\\
\hline
MPSI & 0.00 & 0.07 & 0.11 & 0.15 & 0.20\\
$|f-60|$ & 0.00 & 0.00 & 0.00 & 0.00 & 0.00\\
$V_{error}$ & 0.002 & 0.002 & 0.002 & 0.002 & 0.002\\
MQSI & 0.004 & 0.004 & 0.004 & 0.004 & 0.004\\
\hline
\end{tabular}
\vspace{-4mm}
\end{table}

\color{black}
\subsection{Impact of Reduced Communication and Link Failure}
\begin{figure}
    \centering
    \vspace{-4mm}
    \includegraphics[width=0.5\columnwidth]{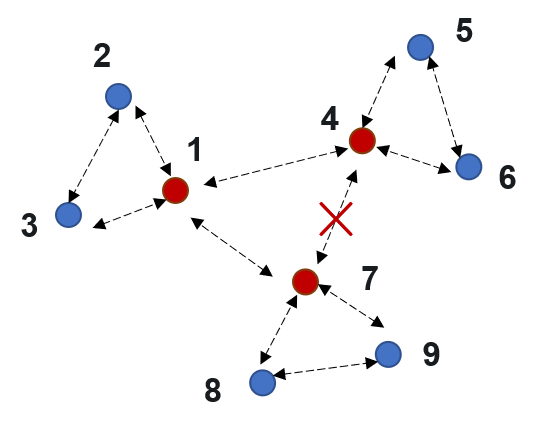}
    \vspace{-2mm}
    \caption{Reduced communication among inverters}
    \label{fig:reduced_comm}
    \vspace{-5mm}
\end{figure}

\begin{figure}
    \centering
      \vspace{-5mm}
    \includegraphics[width=0.95\columnwidth, trim={0mm 0mm 0mm 0mm}, clip] {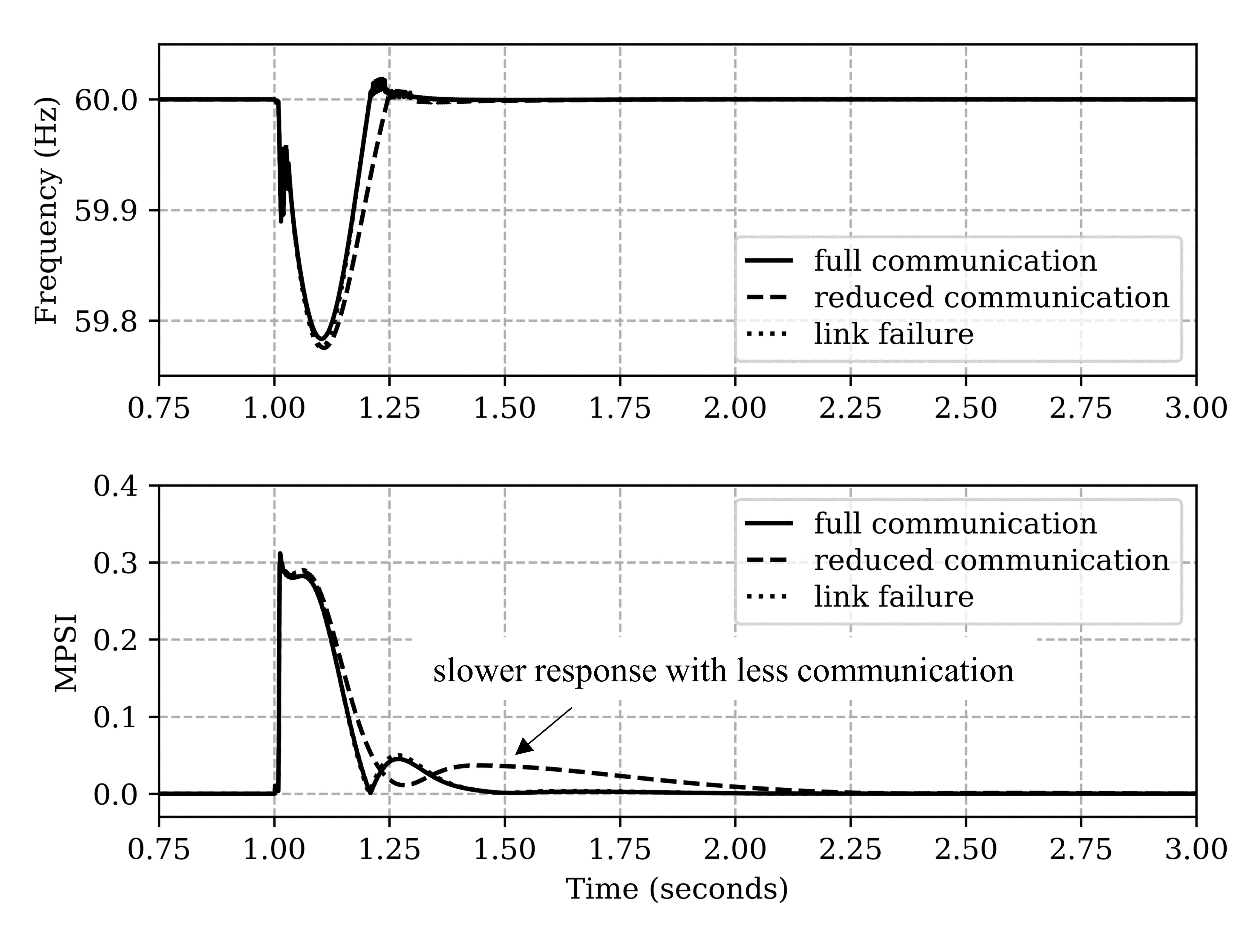}
    \vspace{-4mm}
	\caption{Impact of reduced communication and a communication link failure on (top) frequency response and (bottom) power sharing of the LFC control}
	\vspace{-5mm}
	 \label{fig:reduced_comm_compare}
\end{figure}
\begin{table}
\renewcommand{\arraystretch}{1.15}
\caption{Impact of reduced communication and link failure on controller performance }
\label{tab:reduced_comm_index}
\centering
\begin{tabular}{c c c  c c}
\hline
 cases & $|f-60|$  & MPSI & $V_error$  & MQSI \\
\hline
Full communication & 0.00 & 0.00 & 0.002 & 0.04\\
Reduced communication & 0.00 & 0.00 & 0.002 & 0.14\\
Link failure & 0.00 & 0.00 & 0.002 & 0.05\\
\hline
\end{tabular}
\vspace{-2mm}
\end{table}
After comparative evaluation against existing control strategies, we now test the robustness of the LFC control in cases of reduced communication and communication link failure compared to full communication topology for the islanding disturbance at $t=1$. In  a reduced communication case, each microgrid is considered as one cluster with one leader GFM inverter as shown in \figurename \ref{fig:reduced_comm}. 
The communication across these clusters is only maintained through leader GFMs i.e. inverter 1, 4, and 7 and all other communication across the clusters are disabled. It can be noticed in \figurename \ref{fig:reduced_comm_compare} that even with the reduced communication, the same steady state performance for frequency restoration and power sharing can be achieved as full communication case, though with a little slower control dynamics. Power sharing convergence takes around 0.5 seconds more due to less communication as shown in \figurename \ref{fig:reduced_comm_compare} (bottom). Table \ref{tab:reduced_comm_index} tabulates the impact on all the performance indices. It can be seen that only var sharing is affected by the reduced communication (MQSI increases from 0.04 to 0.14), though it is still significantly better than all other control methods i.e. Case I,II and III. It is worth noting that to achieve a good trade-off between $V_{error}$ and MQSI, scaling parameters $\alpha$ and $\beta$ had to be re-tuned with the reduced communication pattern.

On the other hand, the communication link failure case is simulated by disconnecting a link between inverter 4 and 7 with no update in controller parameters. The overall performance of this case is similar to full communication case as can be seen in \figurename \ref{fig:reduced_comm_compare} and Table \ref{tab:reduced_comm_index} except a slight worsening of var sharing (increase in MQSI from 0.04 to 0.05).

\color{blue}
\subsection{Impact of Communication Topology on Convergence Rate}
\begin{figure}
    \centering
    \includegraphics[width=0.85\columnwidth] {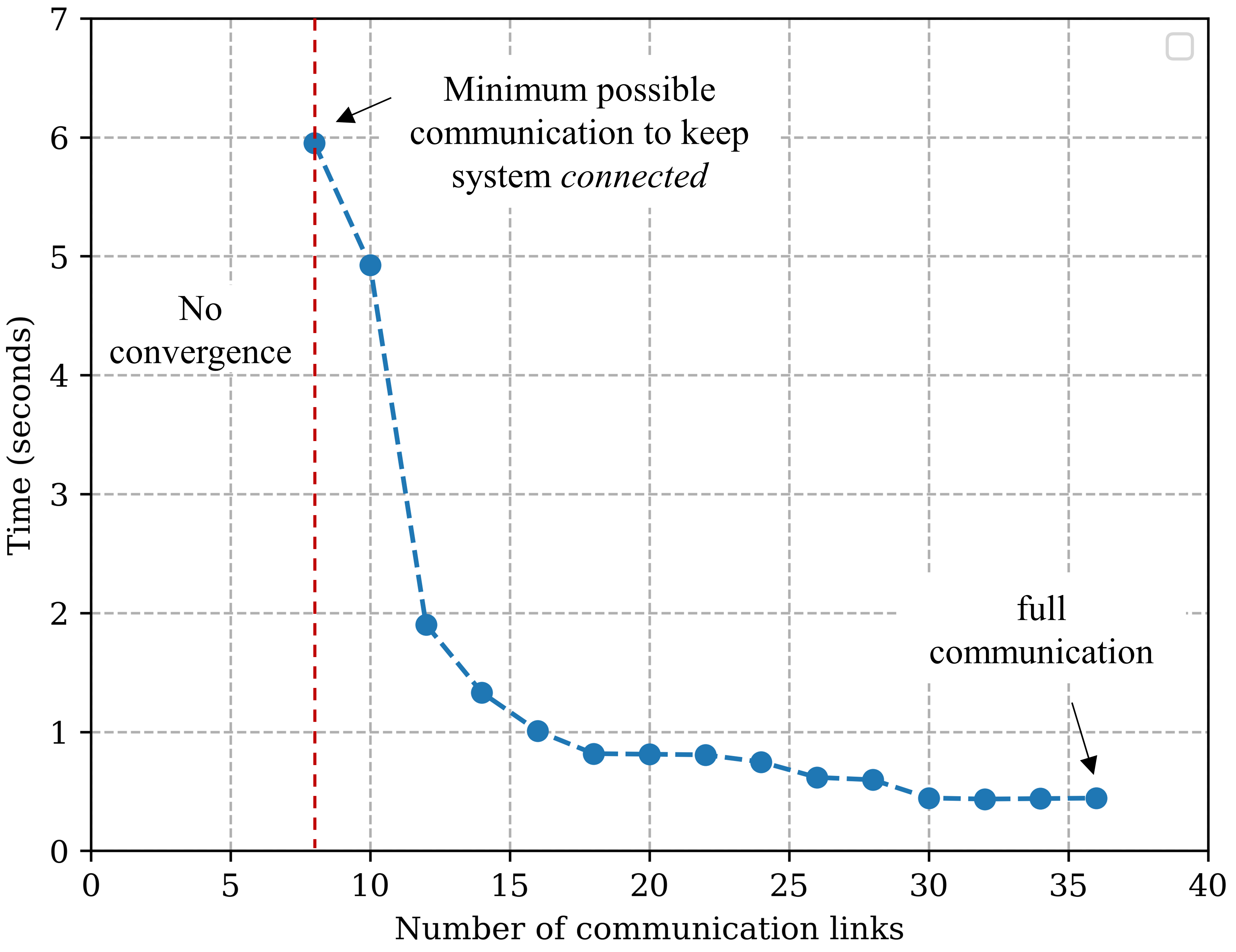}
	\caption{\textcolor{blue}{Impact of communication topology on the convergence rate of the proposed controller. Communication topologies with higher links achieve faster convergence. Minimum links required for consensus convergence, among 9 inverters, is 8.}}
	\vspace{-0mm}
	 \label{fig:link_convergence}
\end{figure}

As discussed earlier, communication graph must be \textit{connected} for convergence of consensus. With $N=9$ in our study, the minimum and maximum possible communication links are 8 and 36, respectively. We conduct a sensitivity study by simulating various connected communication topology with different number of links in the range of 8 to 36, using a random sampling. The consequential convergence rates (seconds required to converge) are plotted in \figurename \ref{fig:link_convergence}. It is observed that compared to the full communication (all 36 links), the convergence rate increases very slowly till topology with around 12 links. However, convergence rate rises sharply closer to the minimum threshold i.e. 8 links. Nonetheless, we observe that the controller converges at optimal steady-state in all topologies. Further, we verify that the consensus is not able to converge the communication links are reduced below 8 as it does not remain connected anymore.
 
\subsection{Performance under Plug-n-Play Case }
\begin{figure}
    \centering
      \vspace{-2mm}
    \includegraphics[width=0.95\columnwidth, clip] {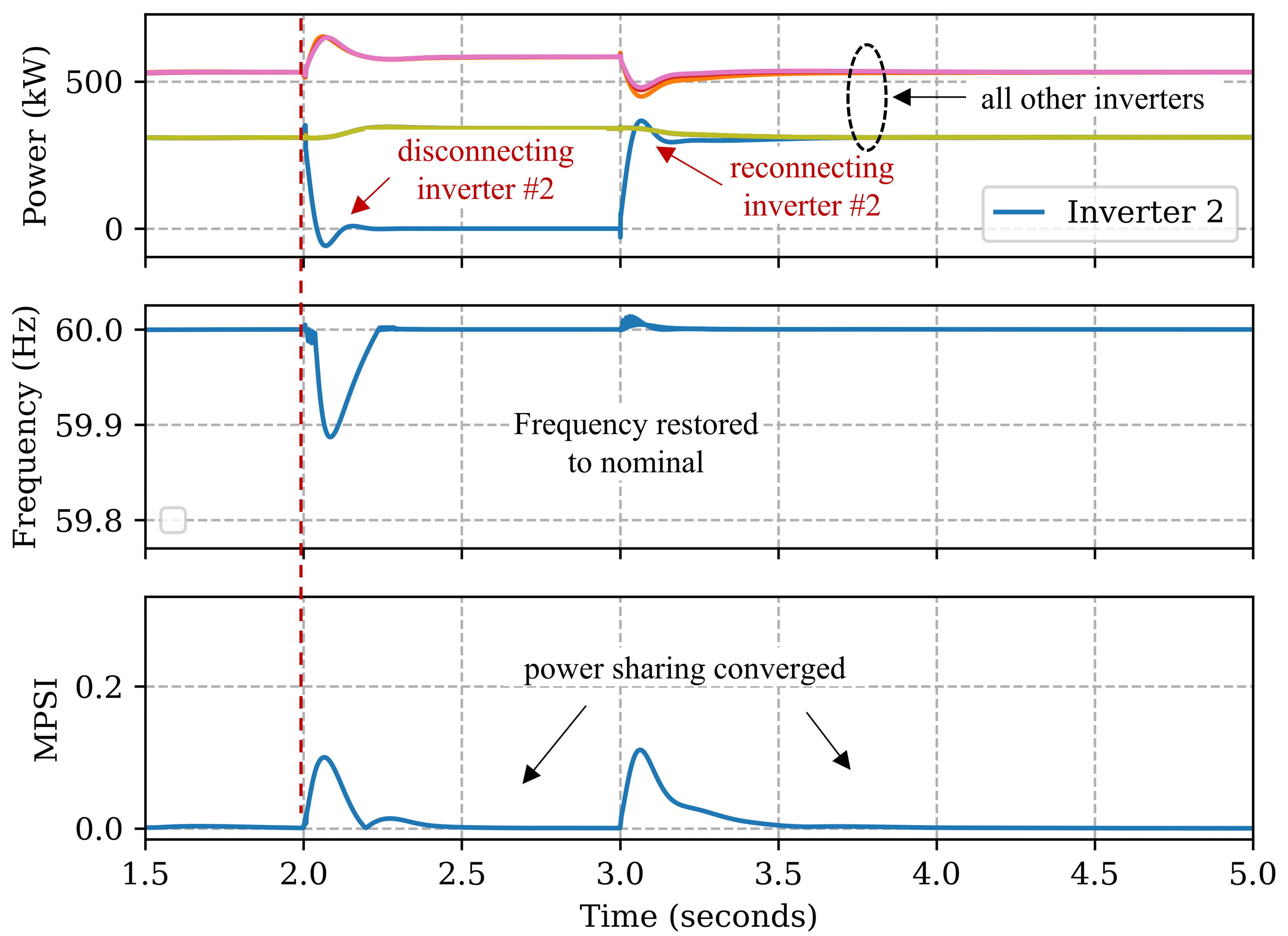}
    \vspace{-2mm}
	\caption{\textcolor{blue}{Controller performance in case of disconnecting and reconnecting inverter 2 to test ‘plug-n-play’ functionality}}
	 \label{fig:plug_n_play}
\end{figure}

To test the ‘plug-n-play’ feature, we simulate a scenario by disconnecting inverter 2 at t=2 and reconnecting back at t=3 as shown in \figurename\ref{fig:plug_n_play}
It can be seen the controller can quickly adapt in both situations and other inverters adjust their power to restore frequency as well converge to equal power sharing. A similar performance is observed at voltage regulation as well. Controller's adaptability with new devices is one of the key factors to facilitate 100\% DER integration in power system.

\color{black}
\subsection{Controller Performance under Microgrid Switching}
\begin{figure}
    \centering
    \includegraphics[width=0.9\columnwidth,trim={1.5mm 8mm 0 0},clip]{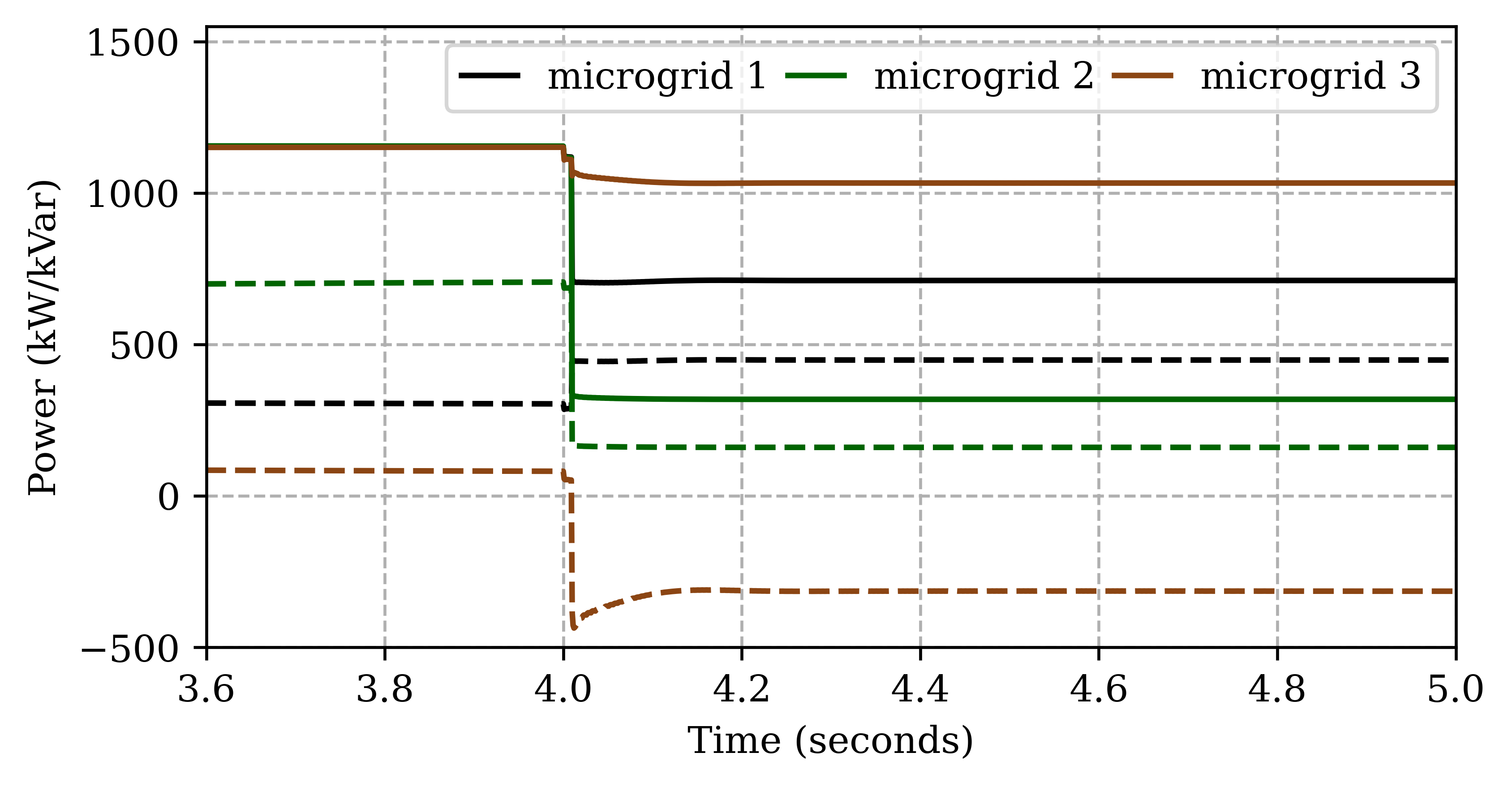}
    \includegraphics[width=0.9\columnwidth,,trim={0 8mm 0 0},clip]{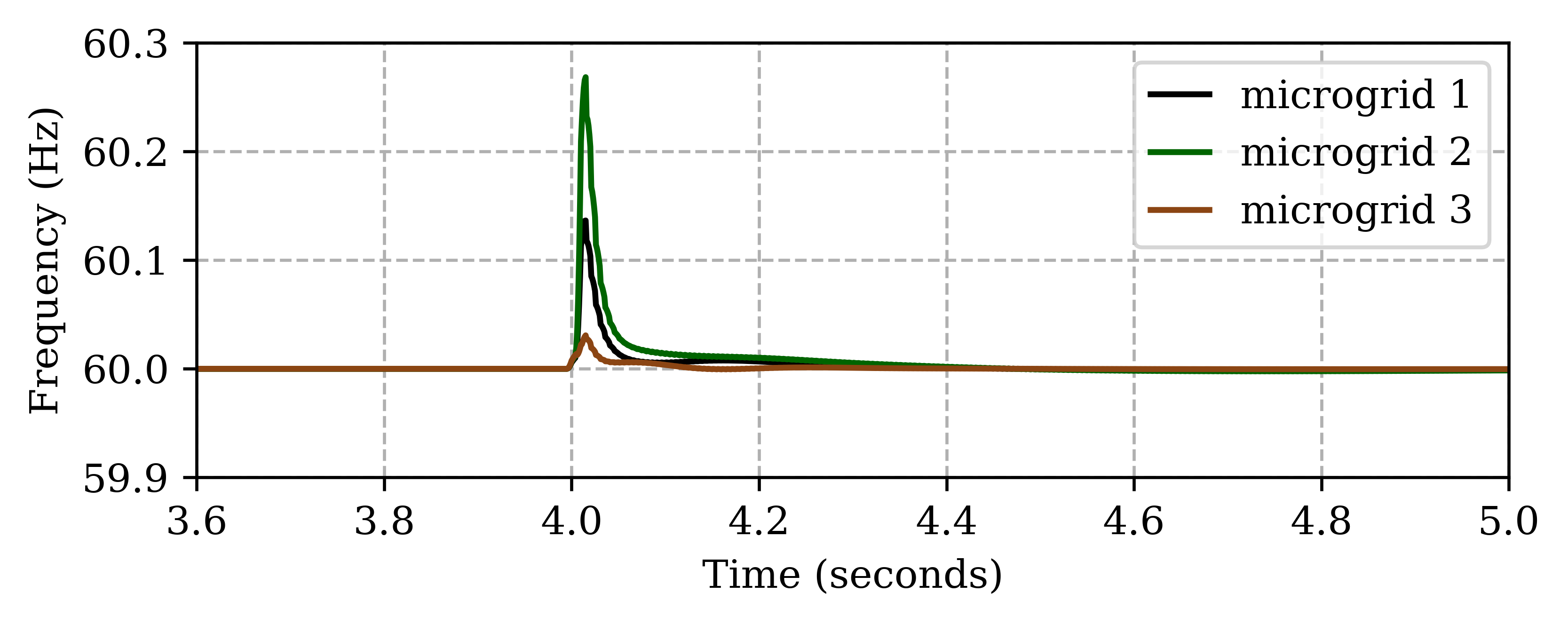}
    \vspace{-4mm}
    \includegraphics[width=0.9\columnwidth]{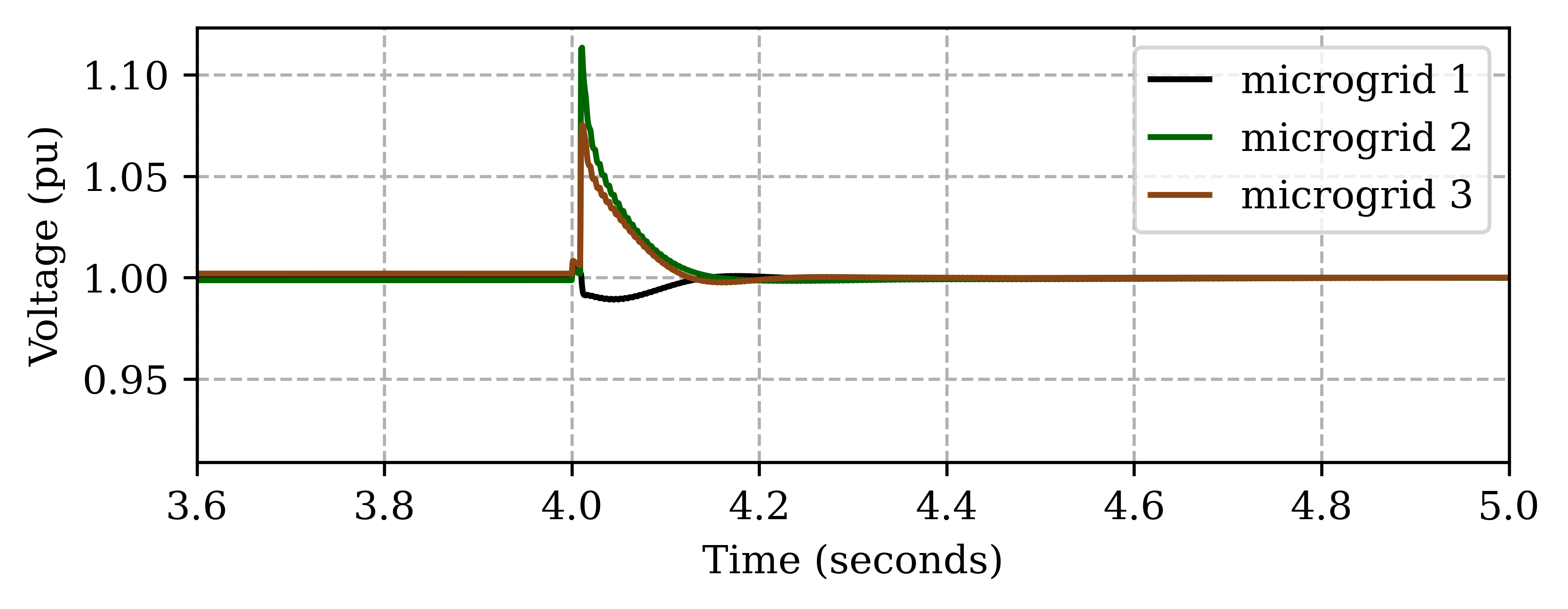}
	\caption{LFC performance in the event of multiple microgrid islanding in a networked microgrid system; (top):total inverter output in each microgrid; (middle): frequency response; (bottom): voltage response of each microgrid}
	\vspace{-4mm}
	 \label{fig:mg_compare}
\end{figure}

We also test the controller performance in the event of multiple microgrids switching in a networked microgrid as shown in \figurename \ref{fig:test_system}. After initial islanding at $t=1$, microgrid switches $sw$ ${1\_ 2}$ and $sw$ ${2 \_ 3}$ are opened at $t=4$ so that all 3 microgrids become islanded networks, each with 1 GFM and 2 GFL inverters. \figurename \ref{fig:mg_compare} (top) shows the total generation profile in each microgrid. The solid and dashed lines denote the real and reactive power respectively. The microgrid 2 can be seen to have the largest adjustment in real power at $t=4$ and therefore it goes through the highest frequency transient peak as shown in \figurename \ref{fig:mg_compare} (middle). Due to different real power adjustments, all 3 microgrids have different transient responses but settle back to nominal frequency. 
\figurename \ref{fig:mg_compare} (bottom) shows the voltage response of each microgrid. Note that since the var demand increases in microgrid 1 (\figurename \ref{fig:mg_compare} (top), dashed black line) unlike microgrid 2 and 3, its voltage transients are also different. Nonetheless, all microgrids are able to settle the voltage at 1 pu.

Table \ref{tab:mg_index} lists all the performance indices values for each microgrid. It is interesting to note that all 3 microgrids are able to achieve an equal var sharing (MQSI=0) along with a accurate voltage regulation ($V_{error}=0$). This is possible in these smaller microgrids because of the electrical proximity of all 3 inverters within each microgrid.
Similarly, an equal real power sharing (MPSI=0) is achieved within each microgrid after switching out from a common power sharing in a large networked-microgrid. The results show the quick adaptability of the proposed LFC controller in an emerging networked microgrid scenarios. 

\begin{table}
\renewcommand{\arraystretch}{1.15}
\caption{The LFC controller performance in an event of multiple microgrid switching }
\label{tab:mg_index}
\centering
\begin{tabular}{c c c  c c}
\hline
&$|f-60|$  & MPSI & $V_error$  & MQSI \\
\hline
microgrid 1 & 0.00 & 0.00 & 0.000 & 0.00\\
microgrid 2 & 0.00 & 0.00 & 0.000 & 0.00\\
microgrid 3 & 0.00 & 0.00 & 0.000 & 0.00\\
\hline
\end{tabular}
\vspace{-4mm}
\end{table}

\vspace{-6mm}

\section{Conclusions}
\vspace{-2mm}
\label{sec.conclusion}
This paper presented a fully coordinated secondary control of GFL and GFM inverters for frequency and voltage regulations   in microgrids. The coordination among inverters was achieved using a leader-follower consensus architecture on both GFM and GFL inverters, where the GFM inverters serve as leaders and GFL inverters serve as followers. The performance of the proposed controller was compared against existing secondary control methods, e.g., uncoordinated and partially coordinated secondary controls, under different disturbances such as islanding, sudden load changes and intermittency. We showed that the coordination is required to obtain equal power sharing among all inverters, while ensuring that the frequency recovers back to the nominal value after the system experiences a disturbance. Similarly, the proposed control improved the voltage regulation while keeping circulating vars within limit. Finally, we showed that this control is robust to communication degradation, intermittency in GFL and can quickly adapt to multiple microgrids switching from a networked microgrid.

%




\ifCLASSOPTIONcaptionsoff
  \newpage
\fi

\vspace{-4mm}
\bibliographystyle{IEEEtran}
\bibliography{pp,ankit_zotero}

\newcommand{\noopsort}[1]{} \newcommand{\printfirst}[2]{#1}
  \newcommand{\singleletter}[1]{#1} \newcommand{\switchargs}[2]{#2#1}
\begin{thebibliography}{10}
\providecommand{\url}[1]{#1}
\csname url@samestyle\endcsname
\providecommand{\newblock}{\relax}
\providecommand{\bibinfo}[2]{#2}
\providecommand{\BIBentrySTDinterwordspacing}{\spaceskip=0pt\relax}
\providecommand{\BIBentryALTinterwordstretchfactor}{4}
\providecommand{\BIBentryALTinterwordspacing}{\spaceskip=\fontdimen2\font plus
\BIBentryALTinterwordstretchfactor\fontdimen3\font minus
  \fontdimen4\font\relax}
\providecommand{\BIBforeignlanguage}[2]{{%
\expandafter\ifx\csname l@#1\endcsname\relax
\typeout{** WARNING: IEEEtran.bst: No hyphenation pattern has been}%
\typeout{** loaded for the language `#1'. Using the pattern for}%
\typeout{** the default language instead.}%
\else
\language=\csname l@#1\endcsname
\fi
#2}}
\providecommand{\BIBdecl}{\relax}
\BIBdecl

\bibitem{Lasseter2001}
B.~{Lasseter}, ``Microgrids [distributed power generation],'' in \emph{2001
  IEEE PES Winter Meeting. Conference Proceedings (Cat. No.01CH37194)}, vol.~1,
  2001, pp. 146--149 vol.1.

\bibitem{TON201284}
D.~T. Ton and M.~A. Smith, ``The u.s. department of energy's microgrid
  initiative,'' \emph{The Electricity Journal}, vol.~25, no.~8, pp. 84 -- 94,
  2012.

\bibitem{Olivares2014}
D.~E. {Olivares}, A.~{Mehrizi-Sani}, A.~H. {Etemadi}, C.~A. {Cañizares},
  R.~{Iravani}, M.~{Kazerani}, A.~H. {Hajimiragha}, O.~{Gomis-Bellmunt},
  M.~{Saeedifard}, R.~{Palma-Behnke}, G.~A. {Jiménez-Estévez}, and N.~D.
  {Hatziargyriou}, ``Trends in microgrid control,'' \emph{IEEE Transactions on
  Smart Grid}, vol.~5, no.~4, pp. 1905--1919, 2014.

\bibitem{wei_circulating-current_2017}
B.~Wei, J.~M. Guerrero, J.~C. Vásquez, and X.~Guo, ``A {Circulating}-{Current}
  {Suppression} {Method} for {Parallel}-{Connected} {Voltage}-{Source}
  {Inverters} {With} {Common} {DC} and {AC} {Buses},'' \emph{IEEE Transactions
  on Industry Applications}, vol.~53, no.~4, pp. 3758--3769, Jul. 2017.

\bibitem{Bidram2012}
A.~{Bidram} and A.~{Davoudi}, ``Hierarchical structure of microgrids control
  system,'' \emph{IEEE Trans. on Smart Grid}, vol.~3, no.~4, pp. 1963--1976,
  2012.

\bibitem{trend_2014_taskforce}
D.~E. Olivares, A.~Mehrizi-Sani, A.~H. Etemadi, C.~A. Cañizares, R.~Iravani,
  M.~Kazerani, A.~H. Hajimiragha, O.~Gomis-Bellmunt, M.~Saeedifard,
  R.~Palma-Behnke, G.~A. Jiménez-Estévez, and N.~D. Hatziargyriou, ``Trends
  in microgrid control,'' \emph{IEEE Transactions on Smart Grid}, vol.~5,
  no.~4, pp. 1905--1919, 2014.

\bibitem{guerrero2008control}
J.~M. Guerrero, L.~Hang, and J.~Uceda, ``Control of distributed uninterruptible
  power supply systems,'' \emph{IEEE Transactions on Industrial Electronics},
  vol.~55, no.~8, pp. 2845--2859, 2008.

\bibitem{prodanovic2006high}
M.~Prodanovic and T.~C. Green, ``High-quality power generation through
  distributed control of a power park microgrid,'' \emph{IEEE Transactions on
  Industrial Electronics}, vol.~53, no.~5, pp. 1471--1482, 2006.

\bibitem{lopes2006defining}
J.~P. Lopes, C.~L. Moreira, and A.~Madureira, ``Defining control strategies for
  microgrids islanded operation,'' \emph{IEEE Transactions on power systems},
  vol.~21, no.~2, pp. 916--924, 2006.

\bibitem{wu20003c}
T.-F. Wu, Y.-K. Chen, and Y.-H. Huang, ``3c strategy for inverters in parallel
  operation achieving an equal current distribution,'' \emph{IEEE Transactions
  on Industrial Electronics}, vol.~47, no.~2, pp. 273--281, 2000.

\bibitem{Brabandere_2007_droop}
K.~De~Brabandere, B.~Bolsens, J.~Van~den Keybus, A.~Woyte, J.~Driesen, and
  R.~Belmans, ``A voltage and frequency droop control method for parallel
  inverters,'' \emph{IEEE Transactions on Power Electronics}, vol.~22, no.~4,
  pp. 1107--1115, 2007.

\bibitem{piagi_autonomous_2006}
P.~Piagi and R.~Lasseter, ``Autonomous control of microgrids,'' in \emph{2006
  {IEEE} {Power} {Engineering} {Society} {General} {Meeting}}, Jun. 2006, pp. 8
  pp.--, iSSN: 1932-5517.

\bibitem{Guerrero_2011_heirarchical}
J.~M. Guerrero, J.~C. Vasquez, J.~Matas, L.~G. de~Vicuna, and M.~Castilla,
  ``Hierarchical control of droop-controlled ac and dc microgrids—a general
  approach toward standardization,'' \emph{IEEE Transactions on Industrial
  Electronics}, vol.~58, no.~1, pp. 158--172, 2011.

\bibitem{savaghebi2012secondary}
M.~Savaghebi, A.~Jalilian, J.~C. Vasquez, and J.~M. Guerrero, ``Secondary
  control scheme for voltage unbalance compensation in an islanded
  droop-controlled microgrid,'' \emph{IEEE Transactions on Smart Grid}, vol.~3,
  no.~2, pp. 797--807, 2012.

\bibitem{Shafiee2014}
Q.~{Shafiee}, J.~M. {Guerrero}, and J.~C. {Vasquez}, ``Distributed secondary
  control for islanded microgrids—a novel approach,'' \emph{IEEE Transactions
  on Power Electronics}, vol.~29, no.~2, pp. 1018--1031, 2014.

\bibitem{Simpson-Porco2015}
J.~W. {Simpson-Porco}, Q.~{Shafiee}, F.~{Dörfler}, J.~C. {Vasquez}, J.~M.
  {Guerrero}, and F.~{Bullo}, ``Secondary frequency and voltage control of
  islanded microgrids via distributed averaging,'' \emph{IEEE Transactions on
  Industrial Electronics}, vol.~62, no.~11, pp. 7025--7038, 2015.

\bibitem{Pattabiraman2018}
D.~{Pattabiraman}, R.~H. {Lasseter.}, and T.~M. {Jahns}, ``Comparison of grid
  following and grid forming control for a high inverter penetration power
  system,'' in \emph{2018 IEEE PESGM}, 2018, pp. 1--5.

\bibitem{Du2020}
W.~{Du}, F.~{Tuffner}, K.~P. {Schneider}, R.~H. {Lasseter}, J.~{Xie},
  Z.~{Chen}, and B.~P. {Bhattarai}, ``Modeling of grid-forming and
  grid-following inverters for dynamic simulation of large-scale distribution
  systems,'' \emph{IEEE Transactions on Power Delivery}, pp. 1--1, 2020.

\bibitem{kenyon_stability_2020}
\BIBentryALTinterwordspacing
R.~W. Kenyon, M.~Bossart, M.~Marković, K.~Doubleday, R.~Matsuda-Dunn,
  S.~Mitova, S.~A. Julien, E.~T. Hale, and B.-M. Hodge,
  ``\BIBforeignlanguage{en}{Stability and control of power systems with high
  penetrations of inverter-based resources: {An} accessible review of current
  knowledge and open questions},'' \emph{\BIBforeignlanguage{en}{Solar
  Energy}}, vol. 210, pp. 149--168, Nov. 2020. [Online]. Available:
  \url{https://www.sciencedirect.com/science/article/pii/S0038092X20305442}
\BIBentrySTDinterwordspacing

\bibitem{hoke_fast_2021}
\BIBentryALTinterwordspacing
A.~O. Hoke, R.~Mahmud, A.~Nelson, D.~Pattabiraman, M.~Asano, D.~Arakawa,
  B.~Pierre, M.~Elkhatib, J.~O. Tan, V.~Gevorgian, C.~Antonio, and E.~Ifuku,
  ``\BIBforeignlanguage{English}{Fast {Grid} {Frequency} {Support} from
  {Distributed} {Energy} {Resources}},'' National Renewable Energy Lab. (NREL),
  Golden, CO (United States), Tech. Rep. NREL/TP-5D00-71156, Mar. 2021.
  [Online]. Available: \url{https://www.osti.gov/biblio/1772978}
\BIBentrySTDinterwordspacing

\bibitem{singhal_real-time_2019}
A.~Singhal, V.~Ajjarapu, J.~Fuller, and J.~Hansen, ``Real-{Time} {Local}
  {Volt}/{Var} {Control} {Under} {External} {Disturbances} {With} {High} {PV}
  {Penetration},'' \emph{IEEE Transactions on Smart Grid}, vol.~10, no.~4, pp.
  3849--3859, Jul. 2019.

\bibitem{johnson_photovoltaic_2016}
J.~Johnson, J.~C. Neely, J.~J. Delhotal, and M.~Lave, ``Photovoltaic
  {Frequency}–{Watt} {Curve} {Design} for {Frequency} {Regulation} and {Fast}
  {Contingency} {Reserves},'' \emph{IEEE Journal of Photovoltaics}, vol.~6,
  no.~6, pp. 1611--1618, Nov. 2016.

\bibitem{pattabiraman_comparison_2018}
D.~Pattabiraman, R.~H. Lasseter., and T.~M. Jahns, ``Comparison of {Grid}
  {Following} and {Grid} {Forming} {Control} for a {High} {Inverter}
  {Penetration} {Power} {System},'' in \emph{2018 {IEEE} {Power} {Energy}
  {Society} {General} {Meeting} ({PESGM})}, Aug. 2018, pp. 1--5, iSSN:
  1944-9933.

\bibitem{poolla_placement_2019}
B.~K. Poolla, D.~Groß, and F.~Dörfler, ``Placement and {Implementation} of
  {Grid}-{Forming} and {Grid}-{Following} {Virtual} {Inertia} and {Fast}
  {Frequency} {Response},'' \emph{IEEE Transactions on Power Systems}, vol.~34,
  no.~4, pp. 3035--3046, Jul. 2019.

\bibitem{awal_unified_2021}
M.~A. Awal and I.~Husain, ``Unified {Virtual} {Oscillator} {Control} for
  {Grid}-{Forming} and {Grid}-{Following} {Converters},'' \emph{IEEE Journal of
  Emerging and Selected Topics in Power Electronics}, vol.~9, no.~4, pp.
  4573--4586, Aug. 2021.

\bibitem{fu_large-signal_2021}
X.~Fu, J.~Sun, M.~Huang, Z.~Tian, H.~Yan, H.~H.-C. Iu, P.~Hu, and X.~Zha,
  ``Large-{Signal} {Stability} of {Grid}-{Forming} and {Grid}-{Following}
  {Controls} in {Voltage} {Source} {Converter}: {A} {Comparative} {Study},''
  \emph{IEEE Transactions on Power Electronics}, vol.~36, no.~7, pp.
  7832--7840, Jul. 2021.

\bibitem{johnson_synchronization_2014}
B.~B. Johnson, S.~V. Dhople, A.~O. Hamadeh, and P.~T. Krein, ``Synchronization
  of {Parallel} {Single}-{Phase} {Inverters} {With} {Virtual} {Oscillator}
  {Control},'' \emph{IEEE Transactions on Power Electronics}, vol.~29, no.~11,
  pp. 6124--6138, Nov. 2014.

\bibitem{zhong_virtual_2016}
Q.-C. Zhong, ``Virtual {Synchronous} {Machines}: {A} unified interface for grid
  integration,'' \emph{IEEE Power Electronics Magazine}, vol.~3, no.~4, pp.
  18--27, Dec. 2016.

\bibitem{chandorkar_control_1993}
M.~Chandorkar, D.~Divan, and R.~Adapa, ``Control of parallel connected
  inverters in standalone {AC} supply systems,'' \emph{IEEE Transactions on
  Industry Applications}, vol.~29, no.~1, pp. 136--143, Jan. 1993.

\bibitem{Olfati-Saber2006}
R.~{Olfati-Saber}, ``Flocking for multi-agent dynamic systems: algorithms and
  theory,'' \emph{IEEE Transactions on Automatic Control}, vol.~51, no.~3, pp.
  401--420, 2006.

\bibitem{Jadbabaie2003}
A.~{Jadbabaie}, {Jie Lin}, and A.~S. {Morse}, ``Coordination of groups of
  mobile autonomous agents using nearest neighbor rules,'' \emph{IEEE
  Transactions on Automatic Control}, vol.~48, no.~6, pp. 988--1001, 2003.

\bibitem{Macellari2017}
L.~{Macellari}, Y.~{Karayiannidis}, and D.~V. {Dimarogonas}, ``Multi-agent
  second order average consensus with prescribed transient behavior,''
  \emph{IEEE Transactions on Automatic Control}, vol.~62, no.~10, pp.
  5282--5288, 2017.

\bibitem{ren_information_2007}
W.~Ren, R.~W. Beard, and E.~M. Atkins, ``Information consensus in multivehicle
  cooperative control,'' \emph{IEEE Control Systems Magazine}, vol.~27, no.~2,
  pp. 71--82, Apr. 2007.

\bibitem{olfati-saber_consensus_2007}
R.~Olfati-Saber, J.~A. Fax, and R.~M. Murray, ``Consensus and {Cooperation} in
  {Networked} {Multi}-{Agent} {Systems},'' \emph{Proceedings of the IEEE},
  vol.~95, no.~1, pp. 215--233, Jan. 2007.

\bibitem{yuan_machine_2020}
H.~Yuan, J.~Tan, Y.~Zhang, S.~Murthy, S.~You, H.~Li, Y.~Su, and Y.~Liu,
  ``Machine {Learning}-{Based} {PV} {Reserve} {Determination} {Strategy} for
  {Frequency} {Control} on the {WECC} {System},'' in \emph{2020 {IEEE} {Power}
  {Energy} {Society} {Innovative} {Smart} {Grid} {Technologies} {Conference}
  ({ISGT})}, Feb. 2020, pp. 1--5, iSSN: 2472-8152.

\bibitem{hoke_rapid_2017}
A.~F. Hoke, M.~Shirazi, S.~Chakraborty, E.~Muljadi, and D.~Maksimovic, ``Rapid
  {Active} {Power} {Control} of {Photovoltaic} {Systems} for {Grid} {Frequency}
  {Support},'' \emph{IEEE Journal of Emerging and Selected Topics in Power
  Electronics}, vol.~5, no.~3, pp. 1154--1163, Sep. 2017.

\bibitem{tuladhar_parallel_1997}
A.~Tuladhar, H.~Jin, T.~Unger, and K.~Mauch, ``Parallel operation of single
  phase inverter modules with no control interconnections,'' in
  \emph{Proceedings of {APEC} 97 - {Applied} {Power} {Electronics}
  {Conference}}, vol.~1, Feb. 1997, pp. 94--100 vol.1.

\bibitem{chassin_gridlab-d_2014}
D.~P. Chassin, J.~C. Fuller, and N.~Djilali,
  ``\BIBforeignlanguage{en}{{GridLAB}-{D}: {An} {Agent}-{Based} {Simulation}
  {Framework} for {Smart} {Grids}},'' \emph{\BIBforeignlanguage{en}{Journal of
  Applied Mathematics}}, vol. 2014, p. e492320, Jun. 2014.

\bibitem{palmintier_design_2017}
B.~Palmintier, D.~Krishnamurthy, P.~Top, S.~Smith, J.~Daily, and J.~Fuller,
  ``Design of the {HELICS} high-performance
  transmission-distribution-communication-market co-simulation framework,'' in
  \emph{2017 {Workshop} on {Modeling} and {Simulation} of {Cyber}-{Physical}
  {Energy} {Systems} ({MSCPES})}, Apr. 2017, pp. 1--6.

\end{thebibliography}
\vspace{-10mm}
\end{document}